\documentclass[english,12pt]{article}
\pdfoutput=1
\usepackage[a4paper,bottom=4.2cm,top=1cm,head=3cm,width=18cm,dvipdfm]{geometry}
\evensidemargin=\oddsidemargin

\usepackage[T1]{fontenc}
\usepackage[latin9]{inputenc}
\usepackage{amstext}
\usepackage{babel}
\usepackage{verbatim}
\usepackage{amsfonts}
\usepackage{mathrsfs}
\usepackage{amsmath}
\usepackage{amssymb}
\usepackage{bm}
\usepackage{extarrows}
\usepackage{slashed}
\usepackage{isodateo}
\usepackage{xcolor}
\usepackage{graphicx}
\usepackage[low-sup]{subdepth}
\usepackage{isodateo}
\usepackage{caption}
\usepackage{subcaption}
\usepackage[linktocpage=true]{hyperref}
\hypersetup{
	colorlinks=true,%
	linkcolor=[rgb]{0,0,1}, 
	citecolor=[rgb]{0,0,1}
}

\numberwithin{equation}{section}

\makeatletter
\@ifundefined{textcolor}{}
{%
	\definecolor{BLACK}{gray}{0}
	\definecolor{WHITE}{gray}{1}
	\definecolor{RED}{rgb}{1,0,0}
	\definecolor{GREEN}{rgb}{0,1,0}
	\definecolor{BLUE}{rgb}{0,0,1}
	\definecolor{CYAN}{cmyk}{1,0,0,0}
	\definecolor{MAGENTA}{cmyk}{0,1,0,0}
	\definecolor{YELLOW}{cmyk}{0,0,1,0}
}

\newcommand{\red}{\color{red}}

\newcommand{\blu}{\color{blue}}

\newcommand{\fr}[2]{\mbox{$\frac{\,{#1}\,}{#2}$}}
\renewcommand{\rm}{\mathrm}
\def\bge{\begin{equation}}
	\def\ede{\end{equation}}
\def\bga{\begin{aligned}}
	\def\eda{\end{aligned}}
\newcommand{\beq}{\begin{equation}}
	\newcommand{\eeq}{\end{equation}}
\newcommand{\bq}{\begin{equation}}
	\newcommand{\eq}{\end{equation}}
\newcommand{\ba}{\begin{array}}
	\newcommand{\ea}{\end{array}}
\newcommand{\beqa}{\begin{eqnarray}}
	\newcommand{\eeqa}{\end{eqnarray}}
\newcommand{\beqs}{\begin{subequations}}
	\newcommand{\eeqs}{\end{subequations}}

\def\nn{\nonumber}

\def\dis{\displaystyle}

\def\({\left(}
\def\){\right)}

\def\End{\end{document}}

\def\d{\text{d}}
\def\ii{{\tt i}}

\def\al{\alpha}
\def\be{\beta}
\def\ga{\gamma}

\def\ep{\epsilon}

\def\ito{\!\rightarrow\!}

\def\OT{\widetilde{\mathcal{O}}}

\renewcommand{\rm}{\mathrm}
\def\bge{\begin{equation}}
\def\ede{\end{equation}}
\def\bga{\begin{aligned}}
\def\eda{\end{aligned}}

\def\nn{\nonumber}

\def\dis{\displaystyle}

\def\({\left(}
\def\){\right)}
\def\[{\left[}
\def\]{\right]}

\def\End{\end{document}}

\def\al{\alpha}
\def\be{\beta}

\def\ga{\gamma}

\def\ep{\epsilon}

\def\si{\Delta}

\def\to{\rightarrow}

\def\ii{\mathrm{i}}

\def\hh{\hat{h}}

\def\cut{\Lambda}

\newcommand{\mT}{\mathcal{T}}

\newcommand{\mL}{\mathcal{L}}
\newcommand{\mO}{\mathcal{O}}

\def\shat{\hat{s}}

\def\OBW{\mathcal{O}_{\widetilde{B}W}^{}}
\def\OGP{\mathcal{O}_{G+}^{}}
\def\OGM{\mathcal{O}_{G-}^{}}
\def\OCP{\mathcal{O}_{C+}^{}}
\def\OCM{\mathcal{O}_{C-}^{}}
\def\TT{\mathcal{T}}
\def\hs{\hspace*{0.3mm}}
\def\hsx{\hspace*{0.5mm}}
\def\hsm{\hspace*{-0.3mm}}
\def\hsmx{\hspace*{-0.5mm}}

\def\to{\rightarrow}
\def\ito{\!\rightarrow\!}

\newlength{\halfpagewidth}
\divide\halfpagewidth by 2

\def\si{\sigma}

\def\End{\end{document}}

\begin{document}
\thispagestyle{empty}

\noindent 

\begin{center}
{\Large\bf Probing Neutral Triple Gauge Couplings
\\[2mm]
with \boldmath{$Z^*\gamma{\hs}(\nu\bar{\nu}\gamma)$} Production at
Hadron Colliders}
\vspace*{8mm}

{{\bf John Ellis}\,$^{a}$,
~~{\bf Hong-Jian He}\,$^{b}$,
~~{\bf Rui-Qing Xiao}\,$^{c}$}

\vspace*{3mm}
{\small 
$^{a}$\,Department of Physics, King's College London, Strand, London WC2R 2LS, UK;\\
Theoretical Physics Department, CERN, CH-1211 Geneva 23, Switzerland;\\
T.\ D.\  Lee Institute, Shanghai Jiao Tong University, Shanghai, China
\\[1.5mm]
$^{b}$\,T.\ D.\ Lee Institute and School of Physics \& Astronomy,\\ 
Key Laboratory for Particle Astrophysics and Cosmology,\\
Shanghai Key Laboratory for Particle Physics and Cosmology,\\
Shanghai Jiao Tong University, Shanghai, China;\\
Physics Department \& Institute of Modern Physics,
Tsinghua University, Beijing, China;\\
Center for High Energy Physics, Peking University, Beijing, China
\\[1.5mm]
$^{c}$\,Department of Physics, King's College London, Strand, London WC2R 2LS, UK;\\
T.\ D.\  Lee Institute and School of Physics \& Astronomy,\\ 
Shanghai Jiao Tong University, Shanghai, China
\\[1.5mm]
({\tt john.ellis@cern.ch}, {\tt hjhe@sjtu.edu.cn}, {\tt xiaoruiqing@sjtu.edu.cn})
}
\end{center}

\vspace*{5mm}
\begin{abstract}
\baselineskip 17pt
\noindent
We study probes of neutral triple gauge couplings (nTGCs) via
$Z^*\ga$ production followed by off-shell decays $Z^*\!\ito\nu\bar{\nu}$  
at the LHC and future $pp$ colliders, 
including both CP-conserving (CPC) and CP-violating (CPV) couplings.\ 
We present the dimension-8 SMEFT operators contributing to nTGCs and
derive the correct form factor formulation for the off-shell vertices  
$Z^*\ga V^*$ ($V\!=\!Z,\ga$) by matching them with the dimension-8 SMEFT operators.\  
Our analysis includes new contributions enhanced by the large off-shell momentum
of $Z^*$, beyond those of the conventional $Z\ga V^*$ vertices 
with on-shell $Z\ga\hs$.\  
We analyze the sensitivity reaches for probing the CPC/CPV nTGC form factors 
and the new physics scales of the dimension-8 nTGC operators at the LHC and 
future 100{\,}TeV $pp$ colliders.\ We compare our new predictions with
the existing LHC measurements of CPC nTGCs in the $\nu\bar\nu\ga$ channel 
and demonstrate the importance of our new method.\
\\[5mm]
KCL-PH-TH/2023-39, CERN-TH-2023-129 
\\[2mm]
Phys.\,Rev.\,D\,(Letter) 108 (2023) L111704, no.11  
{[\,arXiv:{\hs}2308.16887\,]} 
\end{abstract}


\newpage 
\section{\large\hspace{-6.5mm}.\hspace*{1.5mm}Introduction}
\label{sec:1}

Neutral triple gauge couplings (nTGCs) are attracting increased
theoretical and experimental 
interest\,\cite{Ellis:2022zdw}\cite{Ellis:2020ljj}\cite{Ellis:2019zex}\cite{CMS2016nTGC-FF}\cite{Atlas2018nTGC-FF}\cite{nTGC-other}.\ 
This is largely driven by the fact\,\cite{Gounaris:1999kf}\cite{Degrande:2013kka}
that the nTGCs do not appear in the Standard Model (SM) Lagrangian, nor do they show
up in the dimension-6 Lagrangian of the SM Effective Field Theory 
(SMEFT)\,\cite{SMEFT-Rev},
suggesting that they could open up a unique new window to physics beyond the SM
that may first appear at the dimension-8 level.\ 
The SMEFT provides a powerful universal framework to formulate model-independently  
such new physics beyond the SM, 
by parametrizing the low-energy effects of possible high-mass new physics 
in terms of operators composed of the SM fields 
that incorporate the full SU(3)$\otimes$SU(2)$\otimes$U(1)  
gauge symmetry of the SM.\ 

\vspace*{0.5mm}

There have been extensive studies of the dimension-6 
operators\,\cite{SMEFT-Rev}\cite{dim6A}\cite{dim6B}
in the SMEFT, including the experimental constraints
on their coefficients and hence on the associated new physics cutoff scale $\cut{\hs}$.\
But these studies do not involve nTGCs because they first appear as 
a set of dimension-8 operators of the SMEFT.\ 
In general, dimension-8 operators 
make interference contributions to amplitudes at ${O}(1/\Lambda^{4})$,
and thus can contribute to cross sections at the same order.\ But dimension-6
SMEFT operators contributing to amplitudes at ${O}(1/\Lambda^{2})$ 
also contribute to cross sections at ${O}(1/\Lambda^{4})$ in general,
complicating efforts to isolate any dimension-8 contributions.\ The absence of
dimension-6 contributions to nTGCs avoids this complication, 
making them an ideal place to probe 
the new physics at dimension-8 level.

\vspace{1mm}  

Previous phenomenological studies of nTGCs in the SMEFT formalism 
have mainly focused on
CP-conserving (CPC) operators that contribute to scattering amplitudes 
involving the neutral triple gauge vertex (nTGV) 
$Z\ga V^*$ ($V\!\!=\!Z,\ga$) 
through the reaction $f\bar{f}\!\ito Z\ga\,$.\ 
%
In the case of $e^+ e^-$ colliders,
on-shell $Z\ga$ production with 
$Z \!\ito \ell^- \ell^+\!,\hs \nu\bar\nu,\hs q\bar q\hs$ 
final states have been considered, 
but at $pp$ colliders off-shell invisible decays  $Z^*\!\!\ito\!\nu\hs\bar\nu$ 
cannot be separated from on-shell $Z$ decays because of the insufficient kinematic information of $Z$ boson.\   
Hence, for the $\nu\bar{\nu}\ga$ channel it is important to
include the off-shell production of $Z^*\!\ito \nu\bar\nu$ at $pp$ colliders.\

\vspace{0.5mm} 

In this work, we study the nTGVs with two off-shell bosons, 
$Z^*\ga V^*$ ($V\!\! =\!\hsm Z,\ga$),  
which include contributions from additional 
dimension-8 operators that were not considered in the previous nTGC studies.\ 
We include a new analysis of CP-violating (CPV) nTGCs.
As the basis for this work, we first formulate the correct CPC and CPV form factors
of the doubly off-shell nTGVs $Z^*\ga V^*$ 
that are compatible with the full electroweak 
SU(2)$\otimes$U(1) gauge symmetry of the SM\,\cite{foot1}.\
By matching the CPC and CPV dimension-8 nTGC operators with the 
corresponding nTGC form factors, we derive the correct formulations 
of the CPC and CPV $Z^*\ga V^*$ form factors.\ 
We then use these formulations to study the sensitivity reaches of the LHC and the 
projected 100{\,}TeV $pp$ colliders for probing the CPC and CPV nTGCs
through the reaction $pp(q\bar{q})\ito Z^*\ga\ito\nu\bar\nu\ga\hs$.\
We further compare our new predictions with
the existing LHC measurements\,\cite{Atlas2018nTGC-FF} 
of CPC nTGCs in the $\nu\bar\nu\ga$ channel, and demonstrate the importance
of using our new nTGC form factor formulation for 
the correct LHC experimental analysis.

\section{\large\hspace{-6.5mm}.\hspace*{1.5mm}Formulating 
\boldmath{$Z^*\gamma V^*$} Form Factors from Matching the SMEFT}
\label{sec:operators}
\label{sec:2}

In previous works\,\cite{Ellis:2022zdw}\cite{Ellis:2020ljj}\cite{Ellis:2019zex},
we studied the dimension-8 SMEFT operators that generate
nTGCs and their contributions to $Z\ga$ production at the LHC and future colliders.\
In particular, we studied systematically $Z\ga V^*$ 
vertices including their matching with the corresponding SMEFT operators and 
the correct formulation of nTGC form factors that respect the full 
electroweak SU(2)$\otimes$U(1) gauge symmetry of the SM.\ 
However, unlike the case of $e^+ e^-$ collisions 
where the on-shell constraint can be imposed on the invisible decays of  
$Z\!\ito\nu\bar\nu\hs$, this is not possible in $pp$ collisions, for which 
only the missing transverse momentum can be measured.\ 
Moreover, since $Z$ boson is an unstable particle, the invariant-masses of
$\ell^+\ell^-$ and $q\bar{q}$ final states from $Z$ decays are not exactly on-shell, 
in general.\ For these reasons, it is important to study the nTGVs with the
final-state $Z^*$ off-shell, as well as the initial-state $V^*$.

\vspace*{1mm}

The general dimension-8 SMEFT Lagrangian takes the following form:
%
\beqa
\Delta\mathcal{L}(\text{dim-8})
\,=\, \sum_{j}^{}\!
\frac{\tilde{c}_j^{}}{\,\tilde{\cut}^4\,}\mathcal{O}_j^{}
\,=\, \sum_{j}^{}\!
\frac{\,\text{sign}(\tilde{c}_j^{})\,}{\,\cut_j^4\,}\mathcal{O}_j^{}
\,=\, \sum_{j}^{}\!
\frac{1}{\,[\cut_j^4]\,}\mathcal{O}_j^{}
\,,
\label{cj}
\eeqa
where the dimensionless coefficients $\,\tilde{c}_j^{}$ 
may take either sign, 
$\,\text{sign}(\tilde{c}_j^{})\!=\!\pm$\,.\
For each dimension-8 operator $\mathcal{O}_j^{}$\,,
we define in Eq.\eqref{cj} the corresponding effective 
cutoff scale for new physics,
$\,\cut_j^{} \hsmx\equiv\hsm \tilde{\cut}/|\tilde{c}_j^{}|^{1/4}\,$,\
and introduce the notation
$\,[\cut_j^4]\hsm\equiv\hsm\rm{sign}(\tilde{c}_j^{})\cut^4_j\hs$.

\vspace*{1mm}

The following
CPC and CPV nTGC operators include Higgs doublets:
\beqs
\label{eq:dim8H}  
\begin{alignat}{3}
&\text{CPC:} &~~~~ \mO_{\widetilde{B}{W}}^{} &=
	\ii\hs H^\dagger  \widetilde{B}_{\mu\nu}W^{\mu\rho}
	\!\left\{D_{\!\rho}^{},D^\nu\right\}\hsm\!H\!+\!\text{h.c.} \hs, 
\label{eq:obw-CPC} 
\\
\label{eq:obwT-CPC}
&\text{CPC:} &~~~~
\mO_{\widetilde{B W}}^{} &=
\ii\hs H^\dag\hsm \big(\hsm  D_{\!\si}^{}{\widetilde W}^a_{\!\mu\nu}W^{a\mu\si}\!\!+\! D_{\!\si}^{}{\widetilde B}_{\mu\nu}B^{\mu\si}\big)\hsm D^\nu\!H\!+\!\text{h.c.},
\\
&\text{CPV:} &~~~~ \OT_{BW}^{} &=
	\ii\hs H^\dagger  {B}_{\mu\nu}W^{\mu\rho}
	\!\left\{D_{\!\rho}^{},D^\nu\right\}\hsm\!H\!+\!\text{h.c.} \hs,	
\label{eq:obw-CPV} 
\\
&\text{CPV:} &~~~~ \OT_{{W}W}^{} &=
	\ii\hs H^\dagger  {W}_{\mu\nu}W^{\mu\rho}
	\!\left\{D_{\!\rho}^{},D^\nu\right\}\hsm\!H\!+\!\text{h.c.} \hs, 
\label{eq:oww-CPV} 
\\
&\text{CPV:} &~~~~ \OT_{{B}B}^{} &=
	\ii\hs H^\dagger{B}_{\mu\nu}B^{\mu\rho}
	\!\left\{D_{\!\rho}^{},D^\nu\right\}\!\hsm H\!+\!\text{h.c.} \hs, 
\label{eq:obb-CPV}
\end{alignat}
\eeqs
where $H$ denotes the Higgs doublet of the SM, and the covariant derivative term 
$D_{\!\si}^{}{\widetilde B}_{\mu\nu}\!=\hsm 
\partial_{\!\si}^{}{\widetilde B}_{\mu\nu}$
holds for the Abelian field strength.\ 
The operators \eqref{eq:obw-CPC} and \eqref{eq:obw-CPV}-\eqref{eq:obb-CPV}
were given in \cite{Degrande:2013kka}, to which we have further added 
an independent CPC operator \eqref{eq:obwT-CPC}.\ 

\vspace*{1mm}

For the dimension-8 CPC and CPV nTGC operators
containing pure gauge fields only, we have the following:
\\[-6mm]
\beqs
\label{eq:OG+G-} 
\begin{alignat}{3}
\label{eq:OG+}
&\text{CPC:} &~~~~ g \mO_{G+}^{} 
&=\,	
\widetilde{B}_{\!\mu\nu}^{}	 W^{a\mu\rho}
( D_\rho^{} D_\lambda^{} W^{a\nu\lambda} \!+\! D^\nu D^\lambda W^{a}_{\lambda\rho}) \hs,	
\\ 
\label{eq:OG-}
&\text{CPC:} &~~~~ g \mO_{G-}^{} &=\, 		
\widetilde{B}_{\!\mu\nu}^{} W^{a\mu\rho}
( D_\rho^{} D_\lambda^{} W^{a\nu\lambda} \!-\! D^\nu D^\lambda W^{a}_{\lambda\rho}) \hs,
\\
\label{eq:vOG+}
&\text{CPV:} &~~~~ g \OT_{G+}^{} &=\,	
{B}_{\!\mu\nu}^{}	 W^{a\mu\rho}
( D_\rho^{} D_\lambda^{} W^{a\nu\lambda} \!+\! D^\nu D^\lambda W^{a}_{\lambda\rho}) \hs,	
\\ 
\label{eq:vOG-}
&\text{CPV:} &~~~~ g \OT_{G-}^{} &=\, 		
{B}_{\!\mu\nu}^{} W^{a\mu\rho}
( D_\rho^{} D_\lambda^{} W^{a\nu\lambda} \!-\! D^\nu D^\lambda W^{a}_{\lambda\rho}) \hs,
\end{alignat}
\eeqs
where the operators $(\mO_{G+}^{},\hs\mO_{G-}^{})$ are CPC\,\cite{Ellis:2020ljj},  
and the two new CPV operators $(\OT_{G+}^{},\hs\OT_{G-}^{})$ are constructed.\ 
We note that the operators \eqref{eq:dim8H}-\eqref{eq:OG+G-} 
belong to the two classes,\
$F^2\phi^2D^2$ and $F^3D^2$.\  From the classification of \cite{Li:2020gnx}
the relevant dimension-8 operators contains 
$F^2 H^2D^2$, $F^2\psi^2D$, and $F^4$, where $F^2\psi^2D$ and $F^4$ 
do not explicitly contain any nTGC vertices 
and thus are not included here.\
Using integration by parts and equations of motion (EOM), we find that 
$F^3D^2$ type can be converted into three types of operators
$(F^4,\, F^2\psi^2D,\, F^2H^2D^2)$,
where $F^2H^2D^2$ corresponds to our Eq.\eqref{eq:dim8H}.\
Moreover, the $F^2H^2D^2$ type operators contribute to the
form factors $(h_3^V,\,h_1^V)$ only, but not to $(h_4^V,\,h_2^V)$,
as shown by Eqs.\eqref{eq:h-dim8}\eqref{eq:CPV-h12-Cut} below.\ 
Hence, the operator types $F^2\phi^2D^2$ and $F^3D^2$ provide the
optimal basis for the current nTGC study.\
These are further explained in Sec.\,\ref{app:A}
of the Supplemental Material\,\cite{Supp}.\  

\vspace*{1mm}

The conventional formalism for nTGC form factors was proposed 
over 20 years ago\,\cite{Gounaris:1999kf} and respects only
the residual gauge symmetry  $U(1)_{\text{em}}^{}$.\ 
However,  as we stressed in Ref.\,\cite{Ellis:2022zdw},  
it does not respect the full
electroweak SU(2)$\otimes$U(1) gauge symmetry of the SM, 
and leads to large unphysical high-energy behaviors of certain 
scattering amplitudes\,\cite{Ellis:2022zdw}.\ 
We thus proposed\,\cite{Ellis:2022zdw} a new formulation of the CPC form factors 
of the nTGVs $Z\ga V^*$ that is compatible with the full SM gauge group with spontaneous electroweak symmetry breaking.\  We will construct an extended formulation to include the CPV nTGVs for the present study.

\vspace*{1mm}

The doubly off-shell nTGC form factors for $Z^*\ga V^*$ vertices are more complicated than 
the $Z\ga V^*$ form factors.\ 
Matching with the dimension-8 nTGC operators of the SMEFT, 
we can parametrize the $Z^*\ga V^*$ vertices 
in terms of the following form factors: 
\begin{align}
\label{eq:V=Gamma+X1+X3}
V^{\alpha\beta\mu}_{Z^*\ga V^*}
&= \Gamma^{\alpha\beta\mu}_{Z^*\ga V^*}
	+\frac{e}{\,M_Z^2\,}\hs q_1^{\alpha}\hsm X_{1V}^{\beta\mu}
	+\frac{e}{M_Z^2}\hs q_3^\mu\hsm X_{3V}^{\alpha\beta} \,,
\end{align}
where expressions for $X_{1V}^{\beta\mu}$ and $\!X_{3V}^{\alpha\beta}$
are given in Appendix\,\ref{app:A} 
and we find that they make vanishing contributions
to the reaction $f\bar{f}\!\ito Z^{(*)}\ga$ with 
$Z^{(*)}\!\ito\!f'\bar{f}'$.\ 
Hence, the present analysis only involves the vertices
$\Gamma^{\alpha\beta\mu}_{Z^*\ga V^*}$.\ 

\vspace*{1mm}

We present first the CPC parts of the $\Gamma^{\alpha\beta\mu}_{Z^*\ga V^*}$ vertices:
\beqs
\label{eq:Vertex-Z*AV*-CPC}
\begin{align}
\label{eq:Vertex-Z*AA*}
\hspace*{-4mm}
\Gamma_{Z^*\gamma \ga^*}^{\alpha\beta\mu} (q_1^{},q_2^{},q_3^{})
&= \frac{e}{M_Z^2\hs}\hsm\!
\(\hsm\! h_{31}^\ga\!+\!\frac{\hs\hh_3^\ga q_1^2\hs}{M_Z^2}\!\)
\hsmx\! q_3^2q_{2\nu}^{}\epsilon^{\alpha\beta\mu\nu} \!+\!
\frac{e s_W^{}\hh_4q_3^2}{\,2c_W^{}\hsm M_Z^4\hs}
\big(2q_2^\alpha q_{3\nu}^{} q_{2\si}^{} \epsilon^{\beta\mu\nu\si}\hsm\!+\hsm 
q_3^2q_{2\nu}\epsilon^{\alpha\beta\mu\nu}\big) ,
\\
\label{eq:Vertex-Z*AZ*}
\hspace*{-4mm}
\Gamma_{Z^*\gamma Z^*}^{\alpha\beta\mu}  
(q_1^{},q_2^{},q_3^{})
&=	\frac{\,e(q_3^2\!-q_1^2)\,}{M_Z^2}\hsm\!\left[\hsm 
\hat h_3^Zq_{2\nu}^{}\epsilon^{\alpha\beta\mu\nu}
\!+\!\frac{\hat h_4}{\,2M_Z^2\,}
\big(2q_2^\alpha q_{3\nu} q_{2\si} \epsilon^{\beta\mu\nu\si}\hsm\!+\hsm q_3^2q_{2\nu}\epsilon^{\alpha\beta\mu\nu}\big)\!\right]\!\hsm .
\end{align}
\eeqs
In the above, we use the hat symbol to distinguish off-shell form factors 
$(\hat h_3^Z,\hs \hat h_3^\ga,\hs \hat h_4^{})$
from their on-shell counterparts 
$(h_3^\ga,h_3^Z,h_4)$ as studied in \cite{Ellis:2022zdw}.\ 
The CPC form-factor parameters 
$(h_{31}^{\ga},\hs\hh_3^{\ga},\hs \hh_3^{Z},\hs\hh_4^{})$ 
can be mapped precisely to the cutoff scales
$(\cut_{\widetilde{B W}},\hs 
\cut_{\widetilde{B}W}^{},\hs\cut_{G-}^{},\hs\cut_{G+}^{})$
of the dimension-8 operators
$(\mO_{\widetilde{BW}}^{},\hs\mO_{\widetilde{B}W}^{},\hs\mO_{G+}^{},\hs\mO_{G-}^{})$, as follows:
\\[-6mm]
\beqs	
\label{eq:h-dim8}
\begin{align}
\label{eq:hj-Oj}
& \hh_4^{} = \frac{\hat r_4^{}}{\,[\Lambda^4_{G+}]\,},~~~~
\hat h_3^Z = \frac{\hat r_3^Z}{\,[\Lambda^4_{\widetilde{B}W}]\,},~~~~
\hat h_3^\ga =  \frac{\hat r_3^{\ga}}{\,[\Lambda^4_{G-}]\,},~~~~
h_{31}^\ga = \frac{r_{31}^{\ga}}{\,[\cut_{\widetilde{B W}}^4]\,},
\\
&
\hat r_4^{} = -\frac{~v^2\hsm M_Z^2~}{\,s_W^{}c_W^{}\,}\hs,~~~~
\hat r_3^Z=\frac{v^2\hsm M_Z^2}{~2s_W^{}c_W^{}\,} \hs,~~~~
\hat r_3^{\ga} = -\frac{~v^2\hsm M_Z^2~}{~2c_W^2\,}\hs,~~~~
r_{31}^{\ga}= -\frac{\,v^2M_Z^2\,}{\,s_W^{}c_W^{}\,}\hs,
\end{align}
\eeqs
where $\,[\cut_j^4]\!\equiv\hsm\rm{sign}(\tilde{c}_j^{})\cut^4_j\hs$.\
The above relations hold for any momentum $q_1$ of $Z^*$, and 
for $\hs q_1^2\!=\hsm\!M_Z^2\hs$ 
the off-shell form factors $(\hat h_3^Z,\hs \hat h_3^\ga,\hs \hat h_4^{})$ reduce
to the on-shell cases, $\hh_4^{}\!=\!h_4^{}$, $\hh_3^Z\!=\!h_3^Z$, and  $\hh_3^\ga\!+\!h_{31}^{\ga}\!=\!h_3^\ga\hs$,
where $h_{31}^{\ga}$  
is part of the on-shell form factor, 
$\hs h_{31}^{\ga}\!=\!h_3^{\ga}\!-\!\hh_3^{\ga}\hs$, 
for $\hs q_1^2\!=\hsm\!M_Z^2\hs$.\ 

\vspace*{1mm}

We note that the formulations of the off-shell nTGC form factors
$Z^*\ga V^*$ in Eq.\eqref{eq:Vertex-Z*AV*-CPC} (the CPC case) and in 
Eq.\eqref{eq:Vertex-Z*AV*-CPV} (the CPV case) were not given explicitly
in the previous literature,
including Refs.\,\cite{Gounaris:1999kf}-\cite{Degrande:2013kka}.\
As we discuss in Section\,\ref{sec:3}, 
CMS\,\cite{CMS2016nTGC-FF} and ATLAS\,\cite{Atlas2018nTGC-FF} 
measured the CPC nTGC form factors $h_3^{\ga}$ and $h_4^V$ 
via the reaction
$pp(q\bar{q})\ito Z^*\ga\ito\nu\bar\nu\ga\hs$, 
but used the conventional CPC nTGC form factor formulation of $Z\ga V^*$
in which both the final-state bosons $Z\ga$ are assumed to be on-shell\,\cite{Gounaris:1999kf}.\
This means that their measurement of
$h_{3}^{\ga}$ is simply equivalent to measuring the form factor $h_{31}^{\ga}$ 
in Eq.\eqref{eq:Vertex-Z*AA*} above\,\cite{foot2}.\
Hence the CMS and ATLAS analyses\,\cite{CMS2016nTGC-FF}\cite{Atlas2018nTGC-FF} 
missed the new form factor $\hh_{3}^{\ga}$ in Eq.\eqref{eq:Vertex-Z*AA*},
whose contribution dominates over that of $h_{31}^{\ga}$ in the 
$\nu\bar\nu\ga$ channel for both the LHC and 100{\hs}TeV $pp$ colliders, 
as we will demonstrate in Section\,\ref{sec:3}.\  

\vspace*{1mm} 

Next, using the Lagrangian \eqref{eqA:L-CPC+CPV} for nTGVs,
we 
construct the following off-shell CPV nTGVs 
$\Gamma^{\alpha\beta\mu}_{Z^*\ga V^*}$:
\\[-7mm]
\beqs
\label{eq:Vertex-Z*AV*-CPV}
\begin{align}
\hspace*{-3mm}
\Gamma_{Z^*\gamma \ga^*}^{\alpha\beta\mu}(q_1^{},q_2^{},q_3^{})
&=\frac{e}{\,M_Z^2\hsx}\!\!
\(\hsm\! h_{11}^\ga\!+\!\frac{\hs\hh_1^\ga q_1^2\hs}{M_Z^2}\!\)
\hsmx\! q_3^2\big(q_2^{\alpha}g^{\mu\beta}\!\!-\!q_2^\mu g^{\alpha\beta}\big)\hsm 
+\!\frac{\,e\hs s_W^{}\hh_2q_3^2\,}{\,2c_W^{}M_Z^4\,}\hsm 
\big(q_1^2q_2^{\alpha}g^{\mu\beta}\!\!-\!q_3^2q_2^\mu g^{\alpha\beta}\big) \hs ,
\label{eq:Z*AA*-vertex}
\\
\hspace*{-3mm}
\Gamma_{Z^*\gamma Z^*}^{\alpha\beta\mu}(q_1^{},q_2^{},q_3^{})
&=	
\frac{\,e\hs (q_3^2\!-\!q_1^2)\,}{M_Z^2}\!\!\[\hsm \hh_1^Z
\big(q_2^\alpha g^{\mu\beta}\hsm\!-q_2^\mu g^{\alpha\beta}\big)
\!+\!\frac{\,\hh_2\,}{\,2M_Z^2\,}
\big(q_1^2q_2^{\alpha}g^{\mu\beta}\!\!-\!q_3^2q_2^\mu g^{\alpha\beta}\big)\!\]\!,
\label{eq:Z*AZ*-vertex}
\end{align}
\eeqs
which have important differences from the conventional CPV nTGC 
form factors\,\cite{Gounaris:1999kf} as we explain in Appendix\,\ref{app:A}.\
We note that when the final-state $Z$ boson is on-shell ($q_1^2\!=\!M_Z^2$),
the above off-shell form factors should reduce to the on-shell ones,
$h_{11}^{\ga}\!+\!\hh_1^{\ga}\!=\!h_1^{\ga}$,\, $\hh_1^Z\!\!=\!h_1^Z$, 
and $\hh_2^{}\!=\! h_2^{}\hs$.\ In the on-shell limit, 
the longitudinally polarized on-shell external state 
$Z_L^{}(q_1^{})$ should satisfy the equivalence theorem (ET)\,\cite{ET},
which puts nontrivial constraints on the structure of form factors,
as shown in Appendix\,\ref{app:A}.\
 
\vspace*{1mm} 
 
Then, we can match these CPV nTGC form factors to the dimension-8 gauge-invariant CPV
operators \eqref{eq:dim8H}-\eqref{eq:OG+G-} in the broken phase 
and derive the following correspondence relations:
\\[-8mm]
\beqs
	\label{eq:CPV-h12-Cut}
\begin{align}
\hh_1^Z&\,=\, v^2M_Z^2\!\(\hsm\!-\frac{ 1}{\,4[\cut_{WW}^4]\,}
+\frac{c_W^2\!-\hsm s_W^2}{\,4c_Ws_W[\cut_{WB}^4]\,}+   
\frac{1}{\,[\cut_{BB}^4]\,}\!\)\!,
\\
h_{11}^\ga &\,=\, v^2M_Z^2\!\(\!-\frac{s_W }{\,4c_W[\cut_{WW}^4]~}
+\frac{1}{\,2[\cut_{WB}^4]\,}-\frac{c_W}{\,s_W[\cut_{BB}^4]\,}\!\)\!,
\\
\hh_1^\ga &\,=\frac {v^2M_Z^2}{\,4c_W^2[\cut_{\widetilde G-}^4]\,}\hs ,
\\
\hh_2 &\,= -\frac{v^2M_Z^2}{\,2\hs s_W^{}c_W^{}[\cut_{\widetilde G+}^4]\,}\hs .
\end{align}
\eeqs
We find that the Higgs-field-dependent 
CPV operators \eqref{eq:obw-CPV}-\eqref{eq:obb-CPV}  
can generate the form factors 
$(h_{11}^\ga,\hs h_1^Z)$, where $h_{11}^\ga$ is part of the on-shell
form factor $\,h_1^{\ga}=h_{11}^{\ga}\!+\!\hh_1^{\ga}\hs$.

\vspace*{1mm}

In summary, there are eight independent CPC and CPV form factors for 
the nTGC vertices $Z^*\ga V^*$.\ 
These can be mapped to the four CPC operators 
\eqref{eq:obw-CPC}-\eqref{eq:obwT-CPC},\eqref{eq:OG+}-\eqref{eq:OG-} 
and the five CPV operators 
\eqref{eq:obw-CPV}-\eqref{eq:obb-CPV},\eqref{eq:vOG+}-\eqref{eq:vOG-}.\
In the case of $Z\ga V^*$ vertices with on-shell $Z$ and $\ga$, 
there are only six independent parameters 
because $h_{31}^\ga\!+\!\hh_{3}^{\ga}\!=\!h_3^\ga\hs$ and
$h_{11}^{\ga}\!+\!\hh_1^{\ga}\!=\!h_1^{\ga}\hs$.\
Matching the nTGC form factors with the corresponding dimension-8 operators of the SMEFT,
we observe that the $(h_{11}^{\ga},h_{31}^{\ga})$ and 
$(\hh_{1}^Z,\hh_3^Z)$ form factors arise as a consequence of spontaneous 
electroweak symmetry breaking, and would vanish if $\langle H \rangle\!=\!0\,$.\ 
We also note that $\hh_{1,3}^\ga$ and $\hh_{2,4}$ arise from the electroweak rotation of 
the $BW^3W^3$ vertex, so the $\hh_{1}^{\ga}$ term in Eq.\eqref{eq:Z*AA*-vertex}
and $\hh_{3}^\ga$ term in Eq.\eqref{eq:Vertex-Z*AA*}
would vanish if $s_W^{}\!=\hsm 0\,$.\ 

\vspace*{1mm}

In passing, we have also derived the perturbative unitarity bounds on
the cutoff scales $\cut_j^{}$ and the form factors $h_j^V$ 
in Section\,\ref{app:C} of the Supplemental Material.\ We verify that
these bounds are much weaker than our current collider limits
(presented in Section\,\ref{sec:3})
and thus do not affect our collider analyses.

\section{\large\hspace{-6.5mm}.\hspace*{1.5mm}Probing nTGCs with
\boldmath{$Z^*\ga\hs (\nu\bar\nu\ga)$} Production
at Hadron Colliders}
\label{sec:3}
\label{sec:probing}

The invariant mass of $Z^{(*)}\!\!\ito\hsm\!\nu\bar\nu$ decays 
cannot be measured in the reaction 
$pp(q\bar{q})\!\ito\!\nu\bar\nu\ga$ at hadron colliders, so it becomes 
impossible to determine whether the invisible $Z$-decay is on-shell or not.\
Hence analyzing the $\nu\bar\nu\ga$ production with off-shell $Z^*$ decays  
is important for hadron colliders.\

\vspace*{1mm}

The cross section for $q\bar q\to Z^{(*)} \ga$ at a $pp$ collider can be expressed as  
\beqa
\sigma ~=~
\sum_{q,\bar{q}\,}\!\int\!\! \d x_1^{}\hs\d x_2^{} \big[
\mathcal{F}_{\!q/p}^{}(x_1^{},\mu)\hs
\mathcal{F}_{\!\bar{q}/p}^{}(x_2^{},\mu)\hsx
{\sigma}_{\!q\bar q}^{}(\hat s)+(q\leftrightarrow\bar q)\big] \, ,
\eeqa
where the functions $\mathcal{F}_{\!q/p}^{}$ and 
$\mathcal{F}_{\!\bar{q}/p}^{}$ are the parton distribution functions (PDFs) of the quark and antiquark in the proton beams, and 
$\,\hat s\!=\!x_1^{}\hs x_2^{}\hs s\,$.\ 
The PDFs depend on the factorization scale $\mu$, 
which we set to $\,\mu\hsm\! =\hsmx\!\sqrt{\hat{s}\,}/2\,$ 
in our leading-order analysis.\
We use the PDFs of the quarks 
$\hs q\hsm =u,d,s,c,b\hs$ and their antiquarks
determined by the CTEQ collaboration\,\cite{Lai:1999wy}. 
In principle, $\hat s$ can be determined by measuring the invariant-mass of observable final-state particles.\ ATLAS measurements of $M_{\ell\ell\ga}^{}$ 
during the LHC Run-2 reached around 3\,TeV.\ 
Accordingly, in our analysis of $Z^*\ga\hs (\nu\bar\nu\ga)$ production 
we consider the relevant range $\hs\hat s<\!3$\,TeV for the LHC and 
$\hs\hat s\!<\!23$\,TeV for a 100{\hs}TeV $pp$ collider.

\begin{figure}[t]
\centering
\includegraphics[width=8cm,height=6cm]{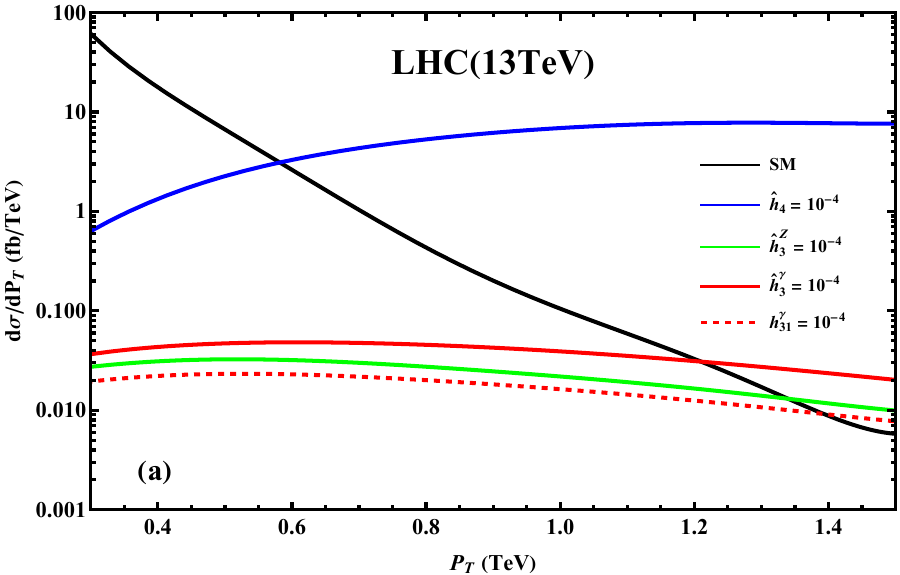}
\hspace*{0.5mm}
\includegraphics[width=8cm,height=6cm]{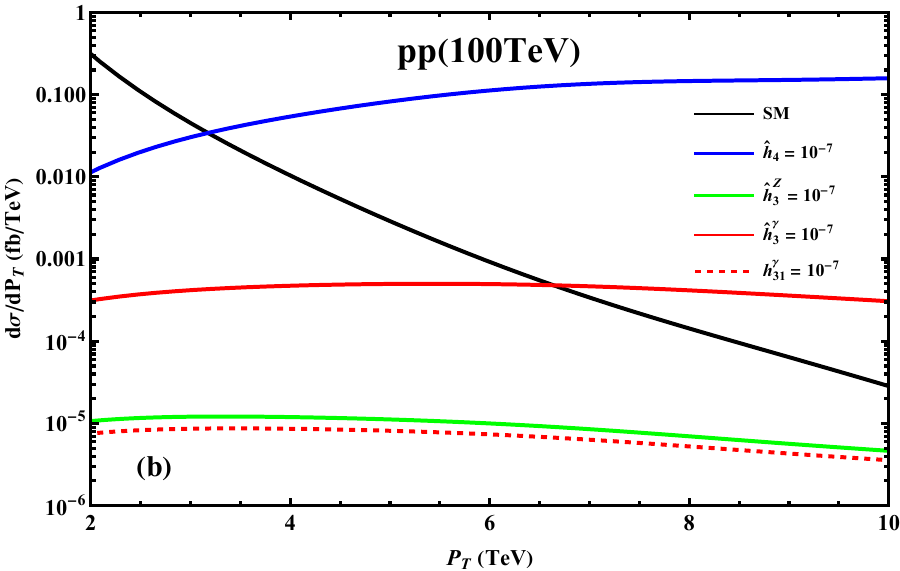}
\vspace*{-2mm}
\caption{\small\hspace*{-2mm}
{\it Differential cross sections for $\nu\bar\nu\ga$ production 
as functions of the photon transverse momentum  $P_T^{\ga}\hs$,
shown in plot\,(a) for the} LHC{\hs}(13{\hs}TeV) {\it and in plot\,(b) 
for the projected} 100{\hs}TeV {\it $pp$ collider.\
In each plot, the SM cross section is given by a black curve, and the 
contributions of the different nTGC form factors (taking a reference value 
$10^{-4}$) are shown by the colored curves.\ 
}}
\label{fig:pt}
\label{fig:1}
\vspace*{1mm}
\end{figure}

\vspace*{1mm}

We compute the partonic cross-section in three parts,
\beq 
\label{eq:sigma=sig0+sig1+sig2}
\sigma(q\bar q\hsm\ito\hsm Z^*\ga )\,=\,
   \sigma_0^{} +\sigma_1^{} +\sigma_2^{}\hs \, ,
\eeq
where $\sigma_0^{}$ is 
the SM contribution, $\sigma_1^{}$ is the contribution 
of the nTGCs interfering with the SM amplitude,
and $\sigma_2^{}$ denotes the squared nTGC contribution.\ 
We present explicit formulae for $(\sigma_0^{},\sigma_1^{},\hs\sigma_2^{})$ 
in Appendix\,\ref{app:sigma-012}.\
The CPC and CPV amplitudes do not interfere 
for $q\bar q\to Z^{(*)}\ga$, hence the contributions of the CPV nTGCs to $\sigma_1^{}$ vanish.\ 
Moreover, we find that $\sigma_1^{}\!\ll\!\sigma_2^{}$ 
holds for the CPC nTGCs at the LHC and future $pp$ colliders, 
making $\sigma_1^{}$ negligible in this analysis\,\cite{foot3}.\ 
We further observe that each CPV nTGC gives the same contribution to
$\sigma_2^{}$ as that of the corresponding CPC nTGC.
 
\vspace*{1mm} 
 
The final state $\gamma$ is the only visible particle when $Z^{(*)}$ has invisible decays, and the longitudinal momentum of the $\nu\bar\nu$ pair cannot be measured at hadron colliders.\ Hence, the photon's transverse momentum $P_T^{\ga}\,$ is the main variable that can be used to distinguish the new physics signals from the SM backgrounds.\ 
We present in Fig.\,\ref{fig:pt} distributions of the $Z^{*}\ga$ cross section 
for the LHC\,(13{\hs}TeV) in plot\,(a) and 
for the projected 100\,TeV $pp$ collider in plot\,(b).\
In each plot, the SM cross section is given by the black curve, and the contributions of 
the nTGC form factors 
$(\hh_4^{},\hh_3^{Z},\hh_3^{\ga},h_{31}^{\ga})$
(taking a reference value $10^{-4}$) 
are shown by the blue, green, red and red-dashed curves.\ 
We observe from Eq.\eqref{eq:Vertex-Z*AA*}
that the $\hh_3^{\ga}$ contribution is strongly enhanced 
by the off-shell $Z^*$ momentum factor $q_1^2/M_Z^2$,
whereas the $\hh_{31}^{\ga}$ contribution is not.\ 
Unlike $\hh_3^{\ga}$,
this does not happen to $\hh_3^Z$ because Eq.\eqref{eq:Vertex-Z*AZ*} 
shows that the numerator factor $q_3^2\hsm - q_1^2$ exhibits a
strong cancellation between  $q_3^2\, (=\!\hat{s})$ and $q_1^2$ 
when $q_1^2$ goes far off-shell,
which is particularly relevant for the high-energy 100{\,}TeV $pp$ collider.\
We find that the $\hh_3^{\ga}$ contribution (red solid curve)
is larger than that of $\hh_{31}^{\ga}$ (red dashed curve) by about a factor of 
$(2\!-\!3)$ for the LHC, as seen in Fig.\,\ref{fig:1}(a), and 
by a large factor of $(43\! -\!77)$ in the case of the 100{\,}TeV $pp$ collider, 
as seen in Fig.\,\ref{fig:1}(b).\  
We note that the $\hh_4^{}$ contribution is much larger than that of 
$\hh_3^Z$, $\hh_3^\ga$, and $h_{31}^\ga$, because the $\hh_4^{}$ terms are enhanced by an extra large momentum factor of
$q_2^{}q_3^{}$ or $q_3^2\,$, as shown in Eq.\eqref{eq:Vertex-Z*AV*-CPC}.\

\vspace*{0.5mm}

In order to optimize the detection sensitivity, we divide events into bins of 
the $P_T^{}(\ga)$ distribution, whose widths we take as
$\Delta P_T^{}\!=\!100$\,GeV for the LHC and 
$\Delta P_T^{}\!=\!500$\,GeV for the 100\,TeV $pp$ collider.\ 
Then, we compute the significance 
$\mathcal{Z}_{\rm{bin}}^{}$ for each bin, and 
construct the following total significance measure:
\beqa
\mathcal{Z}_{\rm{total}}^{}\,=\,
\sqrt{\,\sum\!\mathcal{Z}_{\rm{bin}}^2~} \,.
\label{eq:Zbin}
\eeqa
Since the SM contribution $\sigma_0^{}$ becomes small when the photon $P_T^{}$ 
is high, 
we determine the statistical significance by using the following formula for the 
background-with-signal hypothesis\,\cite{Z0}:
\beq
\label{eq:SS-Z}
\mathcal{Z} \,=\, \sqrt{2\!\(\!B\ln\!
	\frac{\,B\,}{\,B\hsm+\hsm S\,} \!+\!  S\hsm\)\,}
\,=\,\sqrt{2\!\left(\!\sigma_0^{}\hsm\ln\!
	\frac{\,\sigma_0^{}~\,}{\,\sigma_0^{}\hsm+\hsm\Delta\sigma\,} \!+\hsm    \Delta\sigma\!\)\,}\!\times\!\sqrt{\mL\!\times\!\epsilon~},
\eeq
where $\,\mL\,$ is the integrated luminosity and $\epsilon$ denotes the detection efficiency.\ For our analysis we choose an ideal detection efficiency 
$\epsilon\!=\!100\%\hs$ unless specified otherwise.

\begin{table}[t]
\begin{center}
\begin{tabular}{c|ccc||c|ccc}
\hline\hline	
& & & & & &
\\[-4mm]
$\sqrt{s\,}$ & & \hspace*{-12mm}13\,TeV\,
\hspace*{-12mm}
& & & &
&\hspace*{-16mm}100\,TeV\
\\
\hline
& & & & & &
\\[-4.3mm]
$\mL$(ab$^{-1}$) & 0.14 & 0.3 & 3 & 
& 3 & 10 & 30 			
\\
\hline\hline
& & & & & &
\\[-4.3mm]
$|\hat h_{4,2}^{}|\!\times\!10^{6}$
& 11\, & 8.5\, & 4.2\,& $|\hat h_{4,2}^{}|\!\times\!10^{9}$
& 4.5 & 2.9 & 2.0
\\
\hline
& & & & & &
\\[-4.3mm]
$|\hat h_{3,1}^Z|\!\times\!10^{4} $
& 2.2 & 1.7 & 0.90 &$|\hat h_{3,1}^Z|\!\times\!10^{7} $
& 7.0 & 4.8 & 3.4
\\
\hline
& & & & & &
\\[-4.3mm]
$\red |\hat h_{3,1}^\ga|\!\times\!10^{4}$
&\red  1.6 &\red  1.3 &\red  0.67 
&\red  $|\hat h_{3,1}^\ga|\!\times\!10^{7}$
&\red  0.94 &\red  0.62 &\red 0.44\\
\hline
& & & & & &
\\[-4.3mm]
$\blu | h_{31,11}^\ga|\!\times\!10^{4}$
&\blu 2.5 &\blu 2.0 &\blu 1.0 
& $\blu | h_{31,11}^\ga|\!\times\!10^{7}$
&\blu 8.3 &\blu 5.7 &\blu 4.0
\\
\hline\hline
\end{tabular}
\end{center}
\vspace*{-4mm}
\caption{\small\hspace*{-2.5mm}
{\it Sensitivity reaches on probing the CPC and CPV nTGC form factors 
at the $2\hs\sigma$ level, as obtained by analyzing the reaction
$\,p{\hs}p{\hs}(q{\hs}\bar{q})\hsm\ito\hsm Z^*\ga\hsm\ito\hsm \nu\bar{\nu}\ga$		
at the} LHC\,(13\,TeV) {\it and the} 100\,TeV $pp$ 
{\it collider, for the indicated integrated luminosities.\ 
In the last two rows, the $\hh_{3,1}^{\ga}$ sensitivities (red color) 
include the $Z^*$-momentum-square ($q_1^2$) enhanced off-shell effects, 
whereas the $h_{31,11}^{\ga}$ sensitivities (blue color) do not.}}
%
\label{tab:h}
\label{tab:1}
\end{table}

\vspace*{1mm}

We present sensitivities for probing the CPC and CPV nTGC form factors and 
the corresponding new physics scales of the dimension-8 nTGC operators 
in Tables\,\ref{tab:h} and \ref{tab:cut}, respectively. 
For the LHC, we find that the sensitivities to 
$\hs\hh_{2}^{}\hs$ and $\hs\hh_{4}^{}\hs$ can reach the level of 
${O}(10^{-5}\hsm\!-\hsm\!10^{-6})$, 
whereas the sensitivities to $\hh_{3,1}^Z$, $\hh_{3,1}^\ga$ 
and $h_{31,11}^{\ga} $ are of ${O}(10^{-4})$.\ 
The sensitivities for probing the nTGC form factors 
at the projected 100\,TeV $pp$ collider 
are generally much higher than those at the LHC, 
by a factor of $O(10^2\!-\hsm\!10^3)$.\ 
At the LHC the sensitivities to $\hh_{3,1}^\ga$ are stronger than those to
$h_{31,11}^{\ga}$ and $\hh_{3,1}^Z$ by about $(50\!-\!60)\%$, 
whereas at the 100{\hs}TeV $pp$ collider the sensitivities to 
$\hh_{3,1}^{\ga}$ are much higher than those to $h_{31,11}^{\ga}$ 
and $\hh_{3,1}^Z$, by factors of $O(10)\hs$.\

\begin{table}[t]
	\begin{center}
\begin{tabular}{c||ccc|ccc}
\hline\hline	
& & & & & &
\\[-4mm]
$\sqrt{s\,}$ & & \hspace*{-12mm}13\,TeV\,
\hspace*{-12mm}
& & & &\hspace*{-17mm}100\,TeV\
\\
\hline
& & & & & &
\\[-4.3mm]
$\mL$\,(ab$^{-1}$) & 0.14 & 0.3 & 3 &  3 & 10 & 30 			
\\
\hline\hline
& & & & & &
\\[-4.3mm]
$\cut_{G+}${\small (CPC)} 
& 3.2 & 3.5 & 4.1 & 23 & 25 & 28
\\
\hline
& & & & & &
\\[-4.3mm]
$\cut_{G-}^{}${\small (CPC)} 
& 1.2 & 1.3 & 1.5 & 7.7 & 8.5 & 9.3
\\
\hline
& & & & & &
\\[-4.3mm]
$\cut_{\widetilde{B}W}^{}${\small (CPC)} 
& 1.3 & 1.4 & 1.6 & 5.4 & 5.9 & 6.4
\\
\hline
& & & & & &
\\[-4.3mm]
$\cut_{\widetilde{BW}}${\small (CPC)} 
& 1.5 & 1.6 & 1.8 & 6.2 & 6.8 & 7.4
\\
\hline\hline
& & & & & &
\\[-4.3mm]
$\cut_{\widetilde G+}^{}${\small (CPV)} 
& 2.7 & 2.9 & 3.5 & 19 & 21 &23\\
\hline
& & & & & &
\\[-4.3mm]
$\cut_{\widetilde G-}^{}${\small (CPV)} 
& 1.0 & 1.1 & 1.3 & 6.5 & 7.2 & 7.8
\\
\hline
& & & & & &
\\[-4.3mm]
$\cut_{WW}^{}${\small (CPV)}  
& 0.93 & 0.98 & 1.2 & 3.9 & 4.3 & 4.6
\\
\hline
& & & & & &
\\[-4.3mm]
$\cut_{WB}^{}${\small (CPV)}  
& 1.1 & 1.2 & 1.4 & 4.6 & 5.1 &5.5\\
\hline
& & & & & &
\\[-4.3mm]
$\cut_{BB}^{}${\small (CPV)} 
& 1.3 & 1.4 & 1.7 & 5.6& 6.2 & 6.8
\\
\hline\hline
\end{tabular}
\end{center}
\vspace*{-4mm}
\caption{\small\hspace*{-1mm}{\it 
Sensitivity reaches on probing the new physics scales $\cut_j^{}$} (TeV) 
{\it of the dimension-8 nTGC operators
at $2\hs\sigma$ level, as derived by analyzing the reaction
$\,p{\hs}p{\hs}(q{\hs}\bar{q})\!\ito\! Z^*\ga\!\ito\!\nu\bar{\nu}\ga$		
at the} LHC\,(13\,TeV) {\it and at a} 100\,TeV $pp$ 
{\it collider, with the integrated luminosities $\mL$} 
{\it as shown in this table.}}
\label{tab:cut}
\label{tab:2}
\end{table}

\vspace*{1mm}

We recall that  
both CMS (using 19.6/fb of Run-1 data)\,\cite{CMS2016nTGC-FF}
and ATLAS (using 36.9/fb of Run-2 data)\,\cite{Atlas2018nTGC-FF} 
measured the CPC nTGC form factors $(h_3^{V},\,h_4^V)$ via the reaction
$pp(q\bar{q})\ito Z^*\ga\ito\nu\bar\nu\ga\hs$; 
but they used the conventional CPC nTGC form factor formulation of the $Z\ga V^*$
vertex with both $Z\ga$ being on-shell\,\cite{Gounaris:1999kf}.\ 
This differs from our off-shell formulation of the $Z^*\ga V^*$ vertex  
in Eq.\eqref{eq:Vertex-Z*AV*-CPC}, 
which gives the correct nTGC form factor formulation 
for analyzing the reaction
$pp(q\bar{q})\ito Z^*\ga\ito\nu\bar\nu\ga\hs$
at hadron colliders.\ 
For instance, using 36.1/fb of Run-2 data at the LHC{\hs}(13{\hs}TeV), 
ATLAS obtained the following bounds at 95\%\,C.L.{\hs}\cite{Atlas2018nTGC-FF}:
\begin{alignat}{2}
& h_3^\ga\in (-3.7,\,3.7)\!\times\! 10^{-4} \, ,
\hspace*{10mm}
& h_3^Z\in (-3.2,\,3.3)\!\times\! 10^{-4} \, ,
\nn\\[-3mm]
&&
\label{eq:Atlas-h34-exp2018} 
\\[-3.5mm]
& h_4^\ga\in (-4.4,\,4.3)\!\times\! 10^{-7} \, ,
& h_4^Z\in (-4.5,\,4.4)\!\times\! 10^{-7} \, . 
\nn
\end{alignat}
For a quantitative comparison, we impose the same cut 
$P_T^{\ga}\!>\!600$\,GeV 
as that of the ATLAS analysis\,\cite{Atlas2018nTGC-FF}
and choose a detection efficiency $\epsilon\!=\!70\%\hs$  \cite{Atlas2018nTGC-FF}\cite{LiShu}.\ 
We then derive the following LHC bounds on the CPC nTGCs
 at 95\%\,C.L.:
\beq
\label{eq:Atlas-h34-theory}
|h_{31}^\ga|<3.5\!\times\! 10^{-4} \, ,
\hspace*{5mm}
|\hat h_3^\ga|<2.3\!\times\! 10^{-4} \, ,
\hspace*{5mm}
|\hat h_3^Z|< 3.1\!\times\! 10^{-4} \, ,
\hspace*{5mm}
|\hat h_4|< 1.4\!\times\! 10^{-5} \, .
\eeq
Comparing our results \eqref{eq:Atlas-h34-theory} with the ATLAS
result \eqref{eq:Atlas-h34-exp2018}, we find that our bounds on
$h_{31}^{\ga}$ and $\hh_{3}^{Z}$ agree well 
with the ATLAS bounds on $h_{3}^{\ga}$ and $h_{3}^{Z}$
(to within a few percent), 
whereas our bound on $\hh_{3}^{\ga}$ is significantly stronger 
than the ATLAS bound on $h_{3}^{\ga}$ by about 60\%{\hs}.\ 
This is because in our off-shell formulation of Eq.\eqref{eq:Vertex-Z*AV*-CPC}
the form factor  $\hh_{3}^{\ga}$ is enhanced by the $Z^*$ off-shell factor
$q_1^2/M_Z^2$, but $h_{31}^{\ga}$ and $\hh_{3}^{Z}$ are not and thus their
bounds are quite similar to the case of assuming on-shell invisible 
$Z$ decays.\ On the other hand, our $\hh_4^{}$ bound in
Eq.\eqref{eq:Atlas-h34-theory} is much weaker than the ATLAS bounds
on $(h_4^{\ga}, h_4^Z)$, by a large factor of $\sim\hsm\!32{\hs}$,
because the conventional nTGC form factor formulae are incompatible with the
gauge-invariant SMEFT formulation of dimension-8 (including spontaneous 
electroweak gauge symmetry breaking),
as we explain in Appendix\,\ref{app:A} and Table\,\ref{tab:3}.\
We emphasize that the existing ATLAS result\,\cite{Atlas2018nTGC-FF}
used the conventional on-shell nTGC formula 
of the vertex $Z\ga V^*$ \cite{Gounaris:1999kf}  
for analyzing the $\nu\bar{\nu}\ga$ channel, 
which caused a significant underestimate
of the sensitivity to $\hh_3^{\ga}$ by about $60\%{\hs}$.\  
This underlines the importance for the on-going LHC experimental analyses 
to use {\it the correct theoretical formalism} to analyze the
$\nu\bar{\nu}\ga$ channel for probing the nTGCs,
as proposed in this work.

\vspace*{0.5mm}

In Table\,\ref{tab:cut} we demonstrate 
that the sensitivities to new physics scales 
in the coefficients of $\mO_{G+}$ and $\OT_{G+}$ 
can reach $(2.7\!-\hsm 4.1)$\,TeV at the LHC, 
and $(19\hsm -\hsm 28)$\,TeV at the 100\,TeV $pp$ collider.\ 
The sensitivities to probing new physics scales of other nTGC operators are around 
$(1\!-\!1.8){\hs}${\hs}TeV at the LHC and 
$(3.9\!-\!9.3){\hs}$TeV at the 100{\hs}TeV $pp$ collider.\ 
We find that the sensitivities to the coefficients of 
$\mO_{G-}^{}$, $\OT_{G-}^{}$ and $\hh_{1,3}^\ga$
are significantly higher than that of the case by assuming the on-shell  
$Z\ga$ final states\,\cite{Ellis:2022zdw}.
In addition, we have studied the perturbative unitarity constraints\,\cite{unitarity} 
on the nTGC contributions, as presented in Section\,\ref{app:C} of
the Supplemental Material\,\cite{Supp}.\ 
We demonstrate\,\cite{Supp} that the unitarity bounds 
on $\cut_j^{}$ and $h_j^V$ are much weaker than our current collider bounds
in Tables\,\ref{tab:1}-\ref{tab:2}, 
hence they do not affect our collider analyses.

\begin{figure}[t]
\centering
\includegraphics[height=7cm,width=7.5cm]{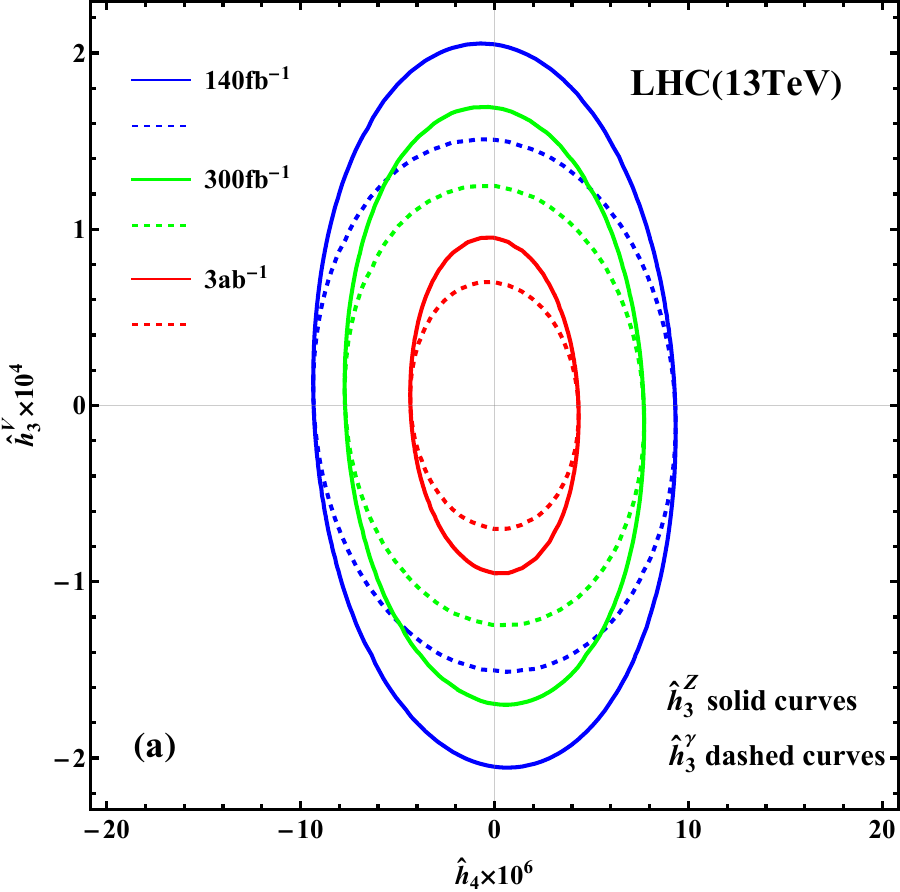}
\hspace*{1mm}
\includegraphics[height=7cm,width=7.5cm]{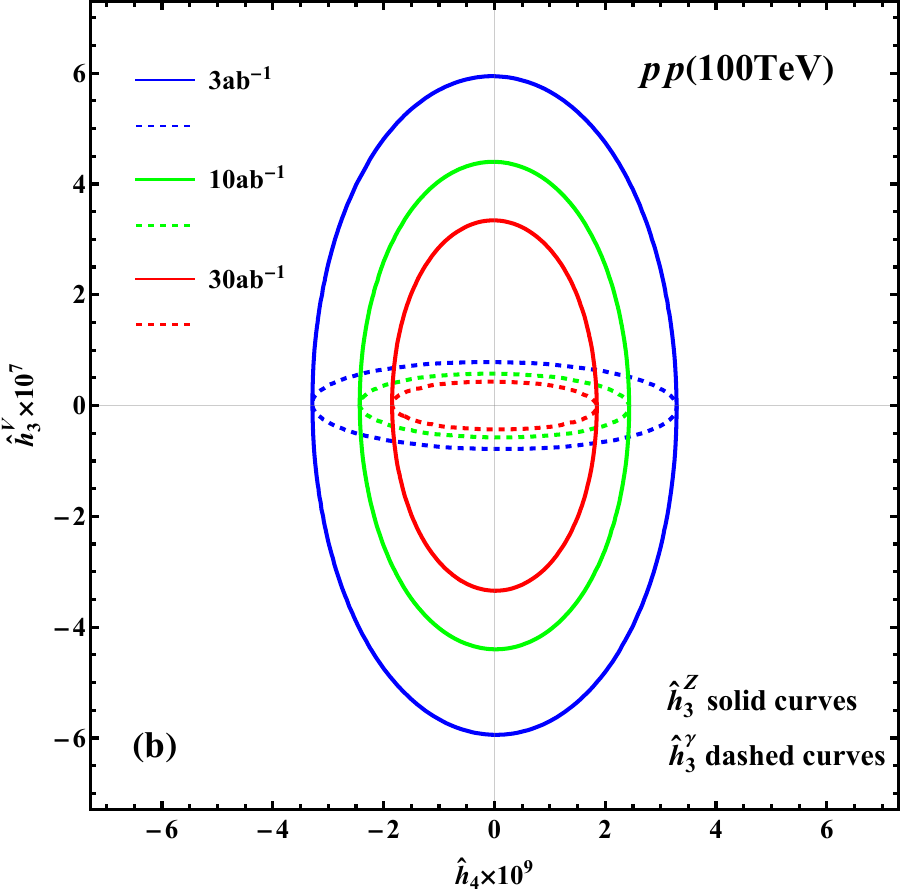}
\\[3mm]
\includegraphics[height=7cm,width=7.5cm]{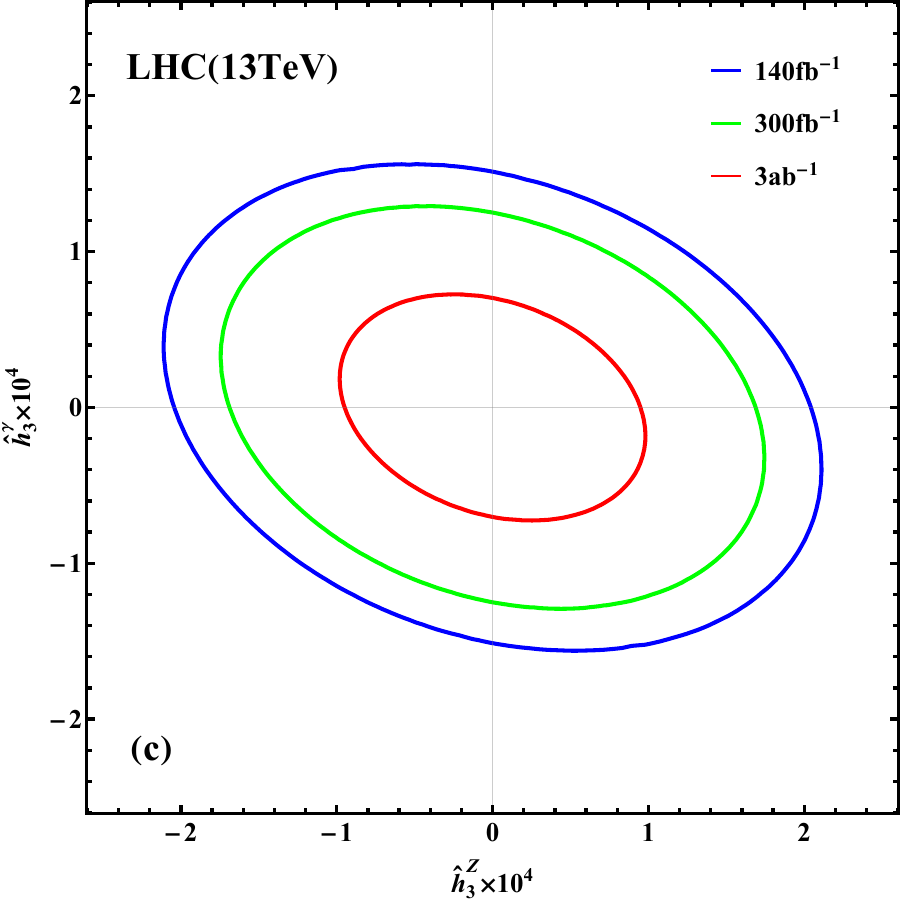}
\hspace*{1mm}
\includegraphics[height=7cm,width=7.5cm]{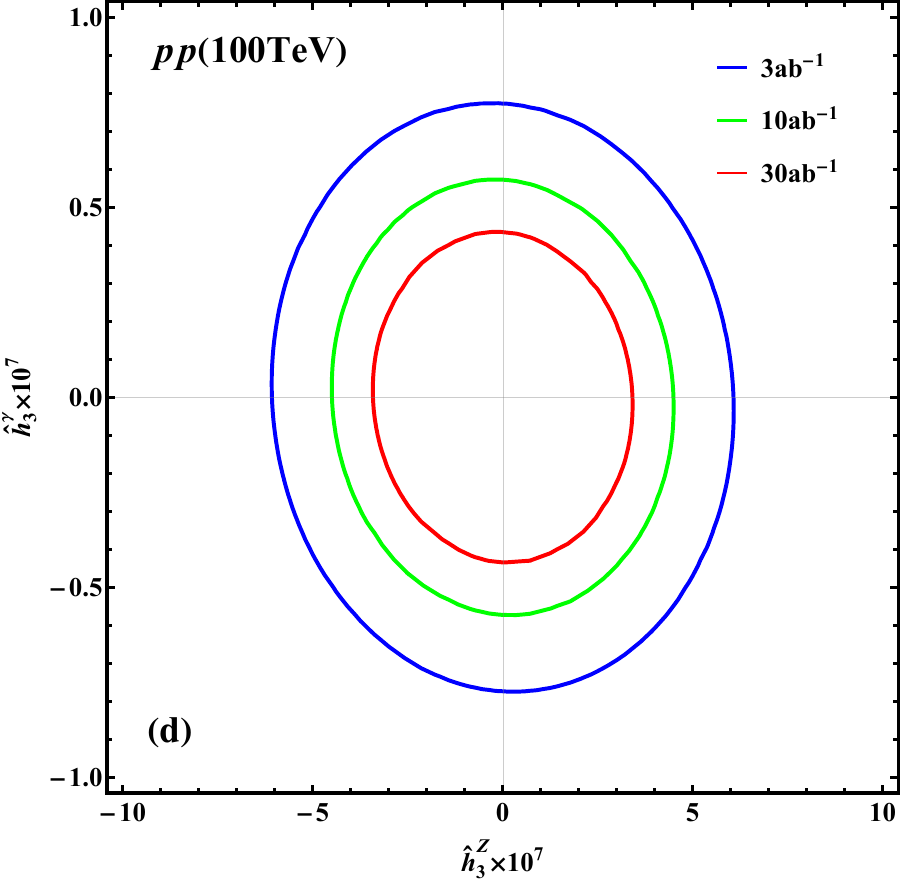}
\vspace*{-2mm}
\caption{\small\hspace*{-2mm}
{\it Correlation contours} (95\%\,C.L.) {\it for the sensitivities to  
each pairs of nTGC form factors at the} LHC{\hs}(13\,TeV) 
[\,panels\,$(a)$ and $(c)$] {\it and the} 100\,TeV {\it $pp$ collider}
[\,panels\,$(b)$ and $(d)$].\
Panels\,(a) and (b) {\it show the correlation contours for $(h_4^{},\,h_3^Z)$ 
(solid curves) and $(h_4^{},\,h_3^{\gamma})$ (dashed curves),
and} panels\,(c) and (d) {\it depict the correlation contours for $(h_3^Z,\,h_3^\gamma)$.\ 
}}
\label{fig:h-corr}
\label{fig:2}
\vspace*{1mm}
\end{figure}

\vspace*{1mm}

Finally, we analyze the correlation contours (95\%\,C.L.) 
between the sensitivities to 
the form factors of the off-shell nTGC vertex $Z^*\ga V^*$, 
as shown in Fig.\,\ref{fig:h-corr}, where we have chosen an ideal detection
efficiency $\ep\!=\!100\%\hs$.\ 
We find that the behavior of the $\hh_4^{}\!-\!\hh_3^Z$ correlation is similar to that of their on-shell counterparts $h_4^{}$ and $h_3^Z$~\cite{Ellis:2022zdw}.\ 
Specifically, the $\hat h_4^{}\!-\!\hh_3^V$ correlation is rather small, 
while the sensitivity to $\hh_3^\ga$ is much greater than that to $h_{31}^\ga$, especially at a 100\,TeV $pp$ collider.\ 
On the other hand, the correlation between $\hh_3^Z\!-\!\hh_3^\ga$ is large at the LHC, 
but almost invisible at the 100\,TeV $pp$ collider.\ 
This is because the $\hh_3^\ga$ contribution is enhanced 
by the off-shell momentum-square $q_1^2/M_Z^2$ of $Z^*$ 
and $\hh_3^Z$ is not, which make the $\hh_3^Z\!-\!\hh_3^\ga$
correlation suppressed by $1/\!\sqrt{q_1^2\,}$.\ At the LHC the on-shell and off-shell
contributions to $h_3^{\ga}$ are comparable and thus the  $\hh_3^Z\hsm -\hsm\hh_3^\ga$
correlation can be significant, whereas their correlation is much suppressed by
$1/\!\sqrt{q_1^2}$ at the 100\,TeV $pp$ collider and thus becomes nearly invisible.

\vspace*{1mm}

We note that no correlation exists between the CPC and CPV nTGCs  
because their amplitudes only differ by $\pm\ii\hs$.\ 
Finally, correlations among $\hh_2^{}$, $\hh_1^Z$, $\hh_1^\ga$, 
and $h_{11}^\ga$ are similar to those of the corresponding CPC nTGCs,
because the squared term $\sigma_2^{}$ dominates the signal cross section
at the LHC and at the 100{\hs}TeV $pp$ collider.

\section{\large\hspace{-6.5mm}.\hspace*{1.5mm}Conclusions}
\label{sec:conx}
\label{sec:4}

In this work, we have demonstrated that the reaction  
$\,p{\hs}p{\hs}(q{\hs}\bar{q})\hsm\ito Z^*\ga\ito\nu\bar{\nu}\ga$
can probe sensitively both the CPC and CPV nTGCs at the LHC
and at the projected 100{\hs}TeV $pp$ colliders.\ 
It has comparable sensitivities to the on-shell production channels 
$\,p{\hs}p{\hs}(q{\hs}\bar{q})\hsm\ito\hsm Z\ga$
with $Z\hsm\ito\ell^+\ell^-\!$ \cite{Ellis:2022zdw}.\
Because the on-shell constraint cannot be imposed on the final state $Z$ boson 
with invisible decays at hadron colliders, 
we studied the $Z^*\ga$ production with off-shell decays  
$Z^{*} \!\hsm\ito\hsm\nu\bar\nu\hs$.\ 
For this, we presented a new general formulation of the CPC/CPV
doubly off-shell $Z^*\ga\hs V^*$ form factors and their nontrivial matching
with the full SU(2)$\otimes$U(1) gauge-invariant SMEFT operators of dimension-8.\ 
In consequence, we found that the conventional on-shell 
form factors for $Z\ga V^*$ vertices\,\cite{Gounaris:1999kf} are inadequate, 
and the conventional form factors $(h_3^V,\hs h_4^V)$ and $(h_1^V,\hs h_2^V)$ 
must be replaced by two sets of new form factors 
$(h_{31}^{\ga},\hs\hh_3^{V},\hs \hh_4^{})$
and $(h_{11}^{\ga},\hs\hh_1^{V},\hs \hh_2^{})$,
as shown in Eqs.\eqref{eq:Vertex-Z*AV*-CPC} and \eqref{eq:Vertex-Z*AV*-CPV}.\ 

\vspace*{1mm}

In the $\nu\bar{\nu}\ga$ production channel at $pp$ colliders,
only the photon $P_T^{\ga}$ is measurable, and  
we have studied how to optimize the sensitivities to nTGCs by proper choices of 
$P_T^{\ga}$ bins at the LHC and the projected 100{\hsx}TeV $pp$ colliders.\ 
Since the interference cross section is always negligible relative to the
squared term at these colliders\,\cite{foot3}, 
the sensitivities to the CPV nTGCs are similar to those to the corresponding CPC nTGCs.\ 

\vspace*{1mm}

We have presented the prospective sensitivity reaches on the 
nTGC form factors $\hh_j^V$ (compatible with the full electroweak gauge symmtry
with spontaneous breaking) at the LHC and the 100\,TeV $pp$ collider 
in Table\,\ref{tab:1}, and the sensitivity reaches on probing the 
new physics scales $\cut_j^{}$ of the corresponding dimension-8 nTGC operators
in Table\,\ref{tab:2}.\
We found in Table\,\ref{tab:1} that
including the off-shell decays $Z^*\!\ito\nu\hs\bar\nu$ 
can increase the sensitivity reaches for $(\hh_3^{\ga},\hs\hh_1^{\ga})$
by $(50\!-\!60)$\% at the LHC
and by a factor of $O(10)$ at the 100{\,}TeV $pp$ collider.\ 
We present in Fig.\,\ref{fig:2} the correlations 
between the sensitivities to each pair of nTGC form factors.

\vspace*{1mm}

We further made quantitative comparisons between the existing 
ATLAS measurements\,\cite{Atlas2018nTGC-FF} of the CPC nTGC form factors
and our new predictions, as shown in 
Eqs.\eqref{eq:Atlas-h34-exp2018}-\eqref{eq:Atlas-h34-theory}.\
We found that for the $\nu\bar\nu\hs\ga$ channel 
the LHC sensitivity reaches on probing the nTGC form factors 
$\hh_3^{\ga}$ and $\hh_4^{}$ 
differ significantly from the ATLAS results (using the conventional 
$Z\ga V^*$ formulae which are inconsistent with spontaneous breaking
of the electroweak gauge symmetry).

\vspace*{1mm}

This work establishes a new perspective for on-going
experimental probes of the new physics in nTGCs 
at the LHC and the projected 100\,TeV $pp$ colliders.\ 
We look forward to continuing the fruitful cooperation with the 
LHC experimental groups, extending their on-going nTGC analyses by 
using our new nTGC formulation,
which consistently incorporates the nTGC form factors with 
the corresponding gauge-invariant dimension-8 operators of the SMEFT.

\vspace*{7mm}
\noindent 
{\bf\large Acknowledgements}
\\[1.5mm]
The work of JE was supported in part by United Kingdom STFC Grant ST/T000759/1, and in
part by the SJTU distinguished visiting fellow programme.\ The work of HJH and RQX was
supported in part by the National NSF of China (under grants 11835005 and 12175136).
RQX has been supported by an International Postdoctoral Exchange Fellowship.

\appendix
\noindent

\vspace*{5mm}

\newpage

\renewcommand{\thesection}{S\arabic{section}}
\renewcommand{\theequation}{S\arabic{equation}}
\renewcommand{\thefigure}{S\arabic{figure}}
\renewcommand{\thetable}{S\arabic{table}}

\begin{center}
	{\Large\bf Probing Neutral Triple Gauge Couplings
		\\[1mm]
		with \boldmath{$Z^*\gamma{\hs}(\nu\bar{\nu}\gamma)$} Production at
		Hadron Colliders
		\\[4mm]
		--- Supplemental Material ---}
	
	\vspace*{8mm}
	
	{{\bf John Ellis}\,$^{a}$,
		~~{\bf Hong-Jian He}\,$^{b}$,
		~~{\bf Rui-Qing Xiao}\,$^{c}$}
	
	\vspace*{3mm}
	{\small 
		$^{a}$\,Department of Physics, King's College London, Strand, London WC2R 2LS, UK;\\
		Theoretical Physics Department, CERN, CH-1211 Geneva 23, Switzerland;\\
		T.\ D.\  Lee Institute, Shanghai Jiao Tong University, Shanghai, China
		\\[1.5mm]
		$^{b}$\,T.\ D.\ Lee Institute and School of Physics \& Astronomy,\\ 
		Key Laboratory for Particle Astrophysics and Cosmology,\\
		Shanghai Key Laboratory for Particle Physics and Cosmology,\\
		Shanghai Jiao Tong University, Shanghai, China;\\
		Physics Department \& Institute of Modern Physics,
		Tsinghua University, Beijing, China;\\
		Center for High Energy Physics, Peking University, Beijing, China
		\\[1.5mm]
		$^{c}$\,Department of Physics, King's College London, Strand, London WC2R 2LS, UK;\\
		T.\ D.\  Lee Institute and School of Physics \& Astronomy,\\ 
		Shanghai Jiao Tong University, Shanghai, China
		\\[1.5mm]
		({\tt john.ellis@cern.ch}, {\tt hjhe@sjtu.edu.cn}, {\tt xiaoruiqing@sjtu.edu.cn})
	}
\end{center}

\vspace*{6mm}

In this Supplemental Material, 
we provide further details to support the analyses 
in the main text\footnote{%
J.~Ellis, H.-J.~He, R.-Q.~Xiao,
Phys.\,Rev.\,D\,(Letter) 108 (2023) L111704, no.11 [arXiv:2308.16887].  
}.\ 
These include three parts.\ In Section\,\ref{app:A}, we first derive systematically 
the neutral triple gauge boson vertices (nTGVs) at the Lagrangian level 
from the corresponding SU(2)$\otimes$U(1) gauge-invariant dimension-8 operators of the 
SM Effective Field Theory (SMEFT) in its electroweak broken phase.\ 
Then,  we present a systematic matching
between the doubly off-shell $Z^*\gamma V^*$ form factors 
and the SMEFT operators, which puts nontrivial constraints on the
neutral triple gauge coupling (nTGC) form factors and ensures 
the correct high energy behaviors.\
In Section\,\ref{app:B}, we provide cross section formulae
for CP-conserving (CPC) and CP-violating (CPV) 
nTGC contributions to the reaction $q\hs\bar{q}\ito Z^*\ga\hs$.\ 
For comparison with the Tables\,I-II of the main text, 
we further present the combined sensitivitiy reaches on probing
the nTGC form factors $h_j^V$ and new physics scales $\cut_j^{}$
for both the $\nu\bar\nu\ga$ channel$^{\blu 1}$ 
and $\ell^-\ell^+\ga$ channel\,\cite{Ellis:2022zdw}.\ 
Finally, in Section\,\ref{app:C}, we present systematically the inelastic 
unitarity bounds on the  nTGC form factors $h_j^V$ and new physics scales 
$\cut_j^{}$ for both the CPC and CPV cases.\ 
These perturbative unitarity bounds are shown 
to be much weaker than our collider sensitivity 
reaches for the relevant energy scales, so they do not affect our present collider analyses.

\section{\large\hspace{-6.5mm}.\hspace*{1.5mm}Matching \boldmath{$Z^*\ga V^*$} Form Factors to
	the SMEFT Operators}
\label{app:A}

For the present nTGC study, we have considered two classes of dimension-8 operators, 
$F^2 H^2D^2$ and $F^3D^2$, where the symbols $(F,\,H,\,D)$ denote the gauge field 
strength, Higgs doublet, and covariant derivative, respectively.\ 
Using integration by parts and equation of motion, 
$F^3D^2$ can be converted to operators of the forms 
$F^4$, $F^2\psi^2D$, and $F^2 H^2D^2$,
where $\psi$ denotes fermion field, and 
the classes of $F^4$ and $F^2\psi^2D$ 
have no direct contribution to nTGCs.\ 
The operator type $F^3D^2$ was not chosen as the basis 
in Ref.\,\cite{Li:2020gnx} since its general classification\,\cite{Li:2020gnx} 
was not optimized for studying any specific type of new physics operators 
such as the nTGCs.\ 
We note that the operator class $F^2 H^2D^2$ 
can contribute only to the form factors $h_1^V$ and $h_3^V$, 
but not to the form factors $h_2^{}$ and $h_4^{}\hs$.\
Thus, we find that without including the class $F^3D^2$,  there would be 
no corresponding operators contributing to the form factors $h_2^{}$ and $h_4^{}$.\ 
Hence, choosing the operator basis of $F^2H^2D^2$ and $F^3D^2$ is necessary 
for a complete formulation of nTGCs and their phenomenological studies.\ 
As we have verified\,\cite{Ellis:2022zdw}\cite{Ellis:2020ljj}, for the reaction 
$f\bar f\to Z\gamma$, the contributions of $h_{2,4}^{}$ from the $F^3D^2$ type
operators are equivalent to that of certain combinations of $F^2H^2D^2$ 
and $F^2\psi^2D$ operators.\ 
We studied\,\cite{Ellis:2022zdw}\cite{Ellis:2020ljj} the fermionic  
operators of type $F^2\psi^2D$\,:
\beqs
\vspace*{-1mm}
\label{eq:nTGC-d8}
\beqa 
\mO_{\!C+}^{} \!\!\!&=&\!\!
\widetilde{B}_{\!\mu\nu}^{}W^{a\mu\rho}\!
\left[D_{\!\rho}^{}(\overline{\psi_{\!L}^{}}T^a\!\gamma^\nu\!\psi_{\!L}^{})
+D^\nu(\overline{\psi_{\!L}^{}}T^a\!\gamma_\rho^{}\psi_{\!L}^{})\right]\!,
\label{eq:OC+}
\\[1.5mm]
\mO_{\!C-}^{} \!\!\!&=&\!\!
\widetilde{B}_{\!\mu\nu}^{}W^{a\mu\rho}\!
\left[D_{\!\rho}^{}(\overline{\psi_{\!L}^{}}T^a\!\gamma^\nu\!\psi_{\!L}^{})
-D^\nu(\overline{\psi_{\!L}^{}}T^a\!\gamma_\rho^{}\psi_{\!L}^{})\right]\!.
\label{eq:OC-}
\end{eqnarray}
\eeqs
They are connected to nTGC operators by equation of motions,
\beqs
\label{eq:OG-EOM}
\begin{eqnarray}
\label{eq:OCP=OGM-OBW}
\OCP \!\!\!&=&\!\! \OGM \!- \OBW \,,
\\[1.5mm]
\label{eq:OCM=OGP-HBW}
\OCM \!\!\!&=&\!\! \OGP -
\{\,\ii H^\dagger\widetilde B_{\mu\nu}{W}^{\mu\rho}\!\left[D_\rho,D^\nu\right]\! H
\!+\ii\,2(D_\rho H)^{\!\dagger}\widetilde B_{\mu\nu}{W}^{\mu\rho}\! D^\nu H+\text{h.c.}\} \hs,
\hspace*{10mm}
\end{eqnarray}
\eeqs
but they do not contribute directly to the nTGCs.

\vspace*{1mm}

Working in the electroweak broken phase of the SM,
SU(2)$_W^{}\otimes$U(1)$_Y^{}\!\to\hs$U(1)$_{\text{em}}$, 
we expand the CPC operators and derive the 
relevant neutral triple gauge vertices (nTGVs) as follows:
{\small 
\beqs
\begin{align}
	\hspace*{-6mm}
	\mO_{G+}^{}(\text{CPC}) 
	&\to\,\frac {e{\hs}v^2}{~4M_Z^2{\hs}s_W^{}c_W^{}\,}  \big(c_W^2\widetilde{A}_{\!\mu\nu}Z^{\mu\rho}\!
	\!-\!s_W^2\widetilde{Z}_{\!\mu\nu} A^{\mu\rho}
	\!+\!c_Ws_W\widetilde{A}_{\!\mu\nu}A^{\mu\rho}
	\!-\!c_Ws_W\widetilde{Z}_{\!\mu\nu}Z^{\mu\rho}\big)
	\partial^2\hsm\Big(\!Z^{\nu}_{\,\,\rho}
	\!+\!\frac{s_W}{c_W}A^{\nu}_{\,\,\rho}\hsm\Big),	
	\\[-1mm]
	\mO_{G-}^{}(\text{CPC}) &\to\,	
	-\frac{e{\hs}v^2}{~4M_Z^2{\hs}s_W^{}c_W^{}\,}
	\big(c_W^2\widetilde{A}_{\!\mu\nu}^{}Z^{\mu\rho}\!-\!s_W^2\widetilde Z_{\!\mu\nu}A^{\mu\rho}+c_Ws_W\widetilde{A}_{\!\mu\nu}A^{\mu\rho}-c_Ws_W\widetilde{Z}_{\!\mu\nu}Z^{\mu\rho}\big)\! \nn\\
	&\quad~~\times	 
	\Big[\partial^2\hsm\Big(\partial^\nu Z_\rho\!+\!\partial_\rho Z^\nu
	\!+\!\frac{s_W}{c_W}\partial^\nu\!A_\rho\!+\!\frac{s_W}{c_W}\partial_\rho A^\nu\Big)
	-2\partial^\nu \partial_\rho\hsm\Big(\partial_\alpha Z^\alpha 
	\!+\!\frac{s_W^{}}{c_W^{}}\partial_\alpha A^\alpha\Big)\hsm\Big] \, ,
\\[0.5mm]
	\mO_{\widetilde{B}W}^{}(\text{CPC}) &\,\to\,
	\frac{e{\hs}v^2}{~4{\hs}s_W^{}c_W^{}\,}	
	\big(c_W^2\widetilde{A}_{\!\mu\nu}^{}Z^{\mu\rho}\!-\hsm s_W^2\widetilde Z_{\!\mu\nu}A^{\mu\rho}+c_Ws_W\widetilde{A}_{\!\mu\nu}A^{\mu\rho}-c_Ws_W\widetilde{Z}_{\!\mu\nu}Z^{\mu\rho}\big)\hsm 	 
	\big(\partial^\nu Z_\rho+\partial_\rho Z^\nu\big)\hs \, ,
	\\[-1mm]
	\mO_{\widetilde{BW}}^{}(\text{CPC}) &\,\to\,
	\frac{-e\hs v^2}{\,2{\hs}s_W^{}c_W^{}\,}	\big(\partial_{\!\rho}^{}
	{\widetilde Z}_{\mu\nu}Z^{\mu\rho}\!+\!\partial_{\!\rho}^{}{\widetilde A}_{\mu\nu}A^{\mu\rho}\big) Z^\nu \, ,
\end{align}
\eeqs
}
\hspace*{-3mm}
where we adopt the notations
$\,\widetilde{V}^{\mu\nu}\!=\epsilon^{\mu\nu\al\be}V_{\al\be}^{}$
and $\,{V}^{\mu\nu}\!=\partial^\mu V^\nu \!-\hsm \partial^\nu V^\mu$
with $V\!=\!A,Z$.\

\vspace*{1mm}

We then expand the CPV operators in the electroweak broken phase 
respecting only the residual SM gauge symmetry 
SU(3)$_C^{}\otimes$U(1)$_{\text{em}}^{}$.\ 
In this way we derive the relevant CPV nTGVs as follows:
\beqs
\label{eq:dim8-brokenP}
{\small 
\begin{align}
	\hspace*{-14mm}
	\OT_{G+}^{}\text{(CPV)}
	&\to\,	\frac {e{\hs}v^2}{\,4M_Z^2{\hs}s_W^{}c_W^{}\,}  A_{\mu\nu}^{}Z^{\mu\rho}\partial^2\Big(Z^{\nu}_{\,\,\,\rho} +
	\frac{\,s_W^{}\,}{c_W^{}}A^{\nu}_{\,\,\,\rho}\Big) \hs ,	
	\hspace*{30mm}
	\\[1.5mm]
	\label{eq:A:TOG-}
	\OT_{G-}^{}(\text{CPV}) 
	&\to\, -\frac{e{\hs}v^2}{\,4M_Z^2{\hs}s_W^{}c_W^{}\,}
	\big[c_{2W}^{}\Big(A_{\!\mu\nu}Z^{\mu\rho}\!+\! {Z}_{\!\mu\nu}A^{\mu\rho}\big)
	+s_{2W}^{}\big({A}_{\!\mu\nu}A^{\mu\rho}\!-\!{Z}_{\!\mu\nu}Z^{\mu\rho}\big)\big]
	\nn\\
	& \hspace*{6mm}
	\times\Big[\partial^2\Big(\partial^\nu Z_\rho+\frac{s_W^{}}{c_W^{}}
	\partial^\nu A_\rho^{}\Big)\!
	-\partial^\nu \partial_\rho\Big(\partial\cdot\! Z
	\hsm +\hsm\frac{s_W^{}}{c_W^{}}\partial\cdot\! A\Big)\Big] \hs ,
\end{align}
\begin{align}
	\OT_{{B}W}^{}(\text{CPV}) &\,\to\,
	\frac{e{\hs}v^2}{\,4s_W^{}c_W^{}\,}
	\big[c_{2W}^{}\big(A_{\!\mu\nu}Z^{\mu\rho}+{Z}_{\mu\nu}^{}A^{\mu\rho})
	\hsm +\hsm s_{2W}^{}\big(A_{\!\mu\nu}A^{\mu\rho}\!-\!{Z}_{\!\mu\nu}Z^{\mu\rho}\big)\big]
	\partial^\nu Z_\rho\hs\,,
	\\[1mm]	
	\OT_{{W}W}^{}(\text{CPV}) &\,\to\,
	-\frac{e{\hs}v^2}{\,4s_W^{}c_W^{}\,}
	\left[s_W^{}c_W^{}\big(A_{\!\mu\nu}Z^{\mu\rho}\!+\!
	Z_{\mu\nu}^{}A^{\mu\rho}\big)+s_{W}^2{A}_{\mu\nu}A^{\mu\rho}\!+\!
	c_W^2{Z}_{\!\mu\nu}Z^{\mu\rho}\right]\!
	\partial^\nu Z_\rho\hs \, ,	
	\\[1mm]	
	\OT_{{B}B}^{}(\text{CPV}) &\,\to\,
	-{e{\hs}v^2} 
	\Big[\!-\!\big({A}_{\!\mu\nu}Z^{\mu\rho}\!+\!Z_{\!\mu\nu}A^{\mu\rho}\big)
	+ \frac{\hs c_W^{}\hs}{s_W^{}}{A}_{\!\mu\nu}A^{\mu\rho}
	+ \frac{\hs s_W^{}\hs}{c_W^{}}{Z}_{\!\mu\nu}Z^{\mu\rho}\Big]
	\partial^\nu Z_\rho\,,
\end{align}
}
\eeqs
\hspace*{-3mm}
where we have defined 
$(s_{2W}^{},\,c_{2W}^{})\!\equiv\!(\sin 2\theta_W^{},\,\cos 2\theta_W^{})$.\ 
We note that in Eq.\eqref{eq:A:TOG-} the factor $\partial\cdot\!Z$ 
vanishes for an on-shell $Z$ boson, while for an off-shell $Z^*$ it also leads to
vanishing results as long as the vector boson couples to a pair of fermions or gauge bosons 
with equal mass.\ The same is true for the factor $\partial\cdot\!A\hs$.\ 
Thus, we can drop terms with either
$\partial\cdot\!Z$ or $\partial\cdot\!A\hs$ 
for the present study.

\vspace*{1mm}

Our next step is to classify the {\it complete structure} 
of the nTGVs in the broken electroweak phase
that are generated by the above SMEFT operators of dimension-8.\  
The Lagrangian of these nTGVs in the electroweak broken phase
contains the following CPC and CPV nTGC form factors,
which generally hold for all fields being off-shell:
\\[-4mm]
{\small 
\beqs
\label{eqA:L-CPC+CPV}
\begin{align}
	\mathcal{L}_{\text{nTGC}}^{\text{CPC}}
	=&\,
	\frac{\,e\hs\hh_3^Z\,}{\,2M_Z^2\,}
	\big(c_W^2\widetilde{A}_{\!\mu\nu}^{}Z^{\mu\rho}
	\!-\hsm s_W^2\widetilde Z_{\!\mu\nu}A^{\mu\rho}+c_Ws_W\widetilde{A}_{\!\mu\nu}A^{\mu\rho}-c_Ws_W\widetilde{Z}_{\!\mu\nu}Z^{\mu\rho}\big)	 
	\big(\partial^\nu Z_\rho\!+\hsm\partial_\rho Z^\nu\big)
	\nn\\
	&\,
	-\!\frac{e\hs\hh_4}{\,2M_Z^4\,}
	\big(c_W^2\widetilde{A}_{\mu\nu}Z^{\mu\rho}\!-\!s_W^2\widetilde{Z}_{\mu\nu} A^{\mu\rho}+c_Ws_W\widetilde{A}_{\!\mu\nu}A^{\mu\rho}-c_Ws_W\widetilde{Z}_{\!\mu\nu}Z^{\mu\rho})\partial^2\Big(\!Z^{\nu}_{\,~\rho}+\!\frac{s_W^{}}{\hs c_W^{}\hs}
	A^{\nu}_{\,~\rho}\hsm\Big)
	\nn\\
	&\,+
	\frac{\,e\hs c_W^{}\hh_3^\ga\,}{\,2s_W^{}M_Z^4\,}
	\big(c_W^2\widetilde{A}_{\mu\nu}^{}Z^{\mu\rho} 
	- s_W^2\widetilde Z_{\!\mu\nu}A^{\mu\rho}+c_Ws_W\widetilde{A}_{\!\mu\nu}A^{\mu\rho}-c_Ws_W\widetilde{Z}_{\!\mu\nu}Z^{\mu\rho}\big)	\times 
	\label{eqA:L-CPC}
	\\ 
	&\,\hspace*{5mm}
	\Big[\partial^2\hsm\Big(\hsm\partial^\nu Z_\rho\!+\hsm\partial_\rho Z^\nu\!+\!
	\frac{s_W^{}}{\hs c_W^{}\hs}\partial^\nu\!A_\rho\!+\!
	\frac{s_W^{}}{\hs c_W^{}\hs}\partial_\rho A^\nu\hsm\Big)\!-\! 
	2\partial^\nu\hsm\partial_\rho
	\Big(\hsm\partial\hsm\cdot\!Z\!+\!\frac{s_W^{}}{\hs c_W^{}\hs}
	\partial\hsm\cdot\!A\Big)\hsm\Big]
	\nn\\
	& +\frac{\,e{\hs}h_{31}^\ga\,}{\,2M_Z^2\,}
	\big(\partial_{\rho}^{}{\widetilde Z}_{\mu\nu}Z^{\mu\rho}
	+\partial_{\rho}^{}{\widetilde A}_{\mu\nu}A^{\mu\rho}\big)Z^\nu \hs ,
	\nn\\[0.5mm]
	\mathcal{L}_{\text{nTGC}}^{\text{CPV}} =&\,
	\frac{\,e\hs \hh_1^Z\,}{\,M_Z^2\,}
	\big(\hsm A_{\mu\nu}Z^{\mu\rho}\!+\!Z_{\mu\nu}A^{\mu\rho}\big)
	\partial^\nu\hsm Z_\rho\!+\!\frac{\,e\hs h_{11}^{\ga}}{\,M_Z^2\,}
	A_{\mu\nu}A^{\mu\rho}\partial^\nu\hsm Z_\rho 
	\!-\!\frac{\,e\hs\hh_2\,}{\,2M_Z^4\,}A_{\mu\nu}Z^{\mu\rho}
	\partial^2\hsm\Big(\!Z^{\nu}_{\,~\rho}\!+\!
	\frac{s_W^{}}{\hs c_W^{}\hs}A^{\nu}_{\,~\rho}\hsm\Big)
	\nn\\
	&\, 
	-\hsm\frac{\,e\hs\hh_1^\ga\,}{\,M_Z^4\,}
	\Big[c_{2W}^{}\big(\hsm A_{\mu\nu}Z^{\mu\rho}\!+\!
	Z_{\mu\nu}A^{\mu\rho}\big) \!+\!s_{2W}^{}
	\big(A_{\!\mu\nu}A^{\mu\rho}\!-\!Z_{\!\mu\nu}Z^{\mu\rho}\big)\Big]
	\label{eqA:L-CPV}
	\\
	&\, \hspace*{8.5mm}
	\times\!\Big[\hsm\partial^2\partial^\nu\hsm
	\Big(\frac{c_W^{}}{s_W^{}} Z_\rho\!+\! A_\rho\Big)
	\hsm -\hsm\partial^\nu\hsm\partial_\rho\partial\hsm\cdot\! 
	\Big(\frac{\hs c_W^{}\hs}{s_W^{}}Z\!+\! A\Big)\!\Big] .
	\nn
\end{align}
\eeqs
}
\hspace*{-3mm}
Matching the above form factor Lagrangian with the corresponding dimension-8 nTGC
operators in Eq.\eqref{eq:dim8-brokenP}, we derive the following relations for the
CPC nTGC form factors:
\beqs	
\label{eqA:h-dim8-CPC}
\begin{align}
	\label{eqA:hj-Oj}
	& \hh_4^{} = \frac{\hat r_4^{}}{\,[\Lambda^4_{G+}]\,} \, ,~~~~
	\hh_3^Z = \frac{\hat r_3^Z}{\,[\Lambda^4_{\widetilde{B}W}]\,},~~~~
	\hh_3^\ga =  \frac{\hat r_3^{\ga}}{\,[\Lambda^4_{G-}]\,} \, ,~~~~
	h_{31}^\ga = \frac{r_{31}^{\ga}}{\,[\cut_{\widetilde{B W}}^4]\,}\,,
	\\
	&
	\hat r_4^{} = -\frac{~v^2\hsm M_Z^2~}{\,s_W^{}c_W^{}\,}\hs,~~~~
	\hat r_3^Z=\frac{v^2\hsm M_Z^2}{~2s_W^{}c_W^{}\,} \hs \, ,~~~~
	\hat r_3^{\ga} = -\frac{~v^2\hsm M_Z^2~}{~2c_W^2\,}\hs \, ,~~~~
	r_{31}^{\ga}= -\frac{\,v^2M_Z^2\,}{\,s_W^{}c_W^{}\,}\hs \, ,
\end{align}
\eeqs
and for the CPV form factors:
\beqs
\label{eqA:h-dim8-CPV}
\begin{align}
\hh_1^Z&\,=\, v^2M_Z^2\!\(\hsm\!-\frac{ 1}{\,4[\cut_{WW}^4]\,}
+\frac{c_W^2\!-\hsm s_W^2}{\,4c_Ws_W[\cut_{WB}^4]\,}+   
\frac{1}{\,[\cut_{BB}^4]\,}\!\)\! \, ,
\\
h_{11}^\ga &\,=\, v^2M_Z^2\!\(\!-\frac{s_W }{\,4c_W[\cut_{WW}^4]~}
+\frac{1}{\,2[\cut_{WB}^4]\,}-\frac{c_W}{\,s_W[\cut_{BB}^4]\,}\!\)\! \, ,
\end{align}
\begin{align}
\hh_1^\ga &\,=\frac {v^2M_Z^2}{\,4c_W^2[\cut_{\widetilde G-}^4]\,}\hs \, ,
\\
\hh_2 &\,= -\frac{v^2M_Z^2}{\,2\hs s_W^{}c_W^{}[\cut_{\widetilde G+}^4]\,}\hs \, ,
\end{align}
\eeqs
where $\,[\cut_j^4]\hsm\equiv\hsm\rm{sign}(\tilde{c}_j^{})\cut^4_j\hs$.\
We stress that the nTGC form factor formulas in Eq.\eqref{eqA:L-CPC+CPV} 
generally hold at the Lagrangian level
{\it with all the fields being off-shell,}
and they are consistently built upon 
the SMEFT broken phase formulation of the nTGCs.\
This provides a fully general and consistent 
off-shell formulation of nTGCs at the Lagrangian level.\ 

\vspace*{1mm}

Next, using the general Lagrangian formulation of nTGVs in Eq.\eqref{eqA:L-CPC+CPV},
we further derive the neutral triple gauge vertices $Z^*\ga V^*$  by requiring one
photon field $A^\mu$ be on-shell,
where the on-shell photon field satisfies the conditions
$\partial^2\!A^\mu\!=\!0\,$ and $\partial_\mu^{} A^\mu\!=\!0\,$.\ 
Thus, we can derive the complete form factor formulation of 
the doubly off-shell nTGVs in momentum space:
\beq
V^{\alpha\beta\mu}_{Z^*\ga V^*} =\,
\Gamma^{\alpha\beta\mu}_{Z^*\ga V^*}
\!+\hsm \frac{e}{M_Z^2\,}X_{1V}^{\beta\mu}q_1^\alpha
+\hsm \frac{e}{M_Z^2\,}X_{3V}^{\alpha\beta}q_3^\mu \hs,
\eeq 
where the expressions of $X_{1V}^{\beta\mu}$ and $\!X_{3V}^{\alpha\beta}$
are given as follows:
{\small 
\beqs
\begin{align}
	\hspace*{-6mm}
	X_{1V}^{\beta\mu} 
	&=\, \epsilon^{\beta\mu\nu\sigma}\hsm q_{2\nu}\hs  q_{3\sigma}\!\hsm 
	\(\!\chi_{10}^V\!+\!\chi_{11}^V\frac{q_1^2}{M_Z^2}\!+\!\chi_{13}^V\frac{q_3^2}{M_Z^2}\!\)
	\!+\!(q_3^2\!-\!q_1^2)\hs g^{\beta\mu}
	\!\hsm\(\!\hsm\tilde\chi_{10}^V\hsmx +\hsmx\tilde\chi_{11}^V\frac{q_1^2}{M_Z^2}
	\hsm +\hsm\tilde\chi_{13}^V\frac{q_3^2}{M_Z^2}\!\)\!,
	\\[0.5mm]
	\hspace*{-6mm}
	X_{3V}^{\alpha\beta}
	&=\, \epsilon^{\alpha\beta\nu\sigma}\hsm q_{2\nu} q_{3\sigma}\!\hsm 
	\(\!\chi_{30}^V\!+\!\chi_{31}^V\frac{q_1^2}{M_Z^2}\!+\!\chi_{33}^V\frac{q_3^2}{M_Z^2}\!\)\!
	+\!(q_3^2\!-\!q_1^2)\hs g^{\alpha\beta}\!
	\(\!\hsm\tilde\chi_{30}^V\hsm +\hsm\tilde\chi_{31}^V\frac{q_1^2}{M_Z^2}
	\!+\hsm\tilde\chi_{33}^V\frac{q_3^2}{M_Z^2}\!\)\!.
\end{align}
\eeqs
}
\hspace*{-3mm}
In the above the additional form factor coefficients $\chi_{ij}^V$ are defined as
follows:
\begin{align}
&\chi_{10}^Z=\frac{\eta_{\widetilde{B}W}^{}}{t_W^{}},\quad \chi_{11}^Z=-\frac{\eta_{G+}^{}}{2t_W^{}}+\frac{t_W^{}\eta_{G-}}{2},
\quad
\chi_{30}^Z=\frac{\eta_{\widetilde{B}W}^{}}{t_W^{}}, \quad \chi_{31}^Z=\frac{\eta_{G+}^{}}{2t_W^{}}-\frac{t_W^{}\eta_{G-}^{}}{2},
\hspace*{15mm}
\nn\\
&\chi_{10}^\ga =\frac{\eta_{\widetilde{B}W}^{}}{2}, \quad 
\chi_{11}^\ga =-\frac{\eta_{G+}^{}}{2}-\frac{\eta_{G-}^{}}{2},\quad
\chi_{30}^\ga =\frac{\eta_{\widetilde{B}W}^{}}{2},\quad 
\chi_{31}^\ga =\frac{\eta_{G+}}{2}-\frac{t_{W}^2\eta_{G-}^{}}2,
\nn\\
&\tilde\chi_{10}^Z =-\frac{\eta_{{B}W}^{}}{2t_{2W}^{}}
+\frac{\eta_{WW}^{}}4-\eta_{{B}B}^{}, \quad 
\tilde\chi_{11}^Z =\frac{\eta_{\widetilde G+}^{}}{4s_{2W}^{}},
\quad
\chi_{13}^Z=-\frac{\eta_{\widetilde G+}^{}}{4s_{2W}^{}},
\nn\\
&\tilde\chi_{30}^Z=\frac{\eta_{{B}W}^{}}{2t_{2W}^{}}
-\frac{\eta_{WW}^{}}{4}+\eta_{{B}B}^{},
\quad 
\tilde\chi_{31}^Z=\frac{\eta_{\widetilde G+}^{}}{4s_{2W}^{}},
\quad
\chi_{33}^Z=-\frac{\eta_{\widetilde G+}^{}}{4s_{2W}^{}},
\\
&\tilde\chi_{10}^\ga =-\frac{\eta_{{B}W}^{}}4 +\frac{t_W^{}\eta_{{W}W}^{}}{8}+\frac{\eta_{{B}B}^{}}{2t_W^{}},
\quad 
\tilde\chi_{11}^\ga =\frac{\eta_{\widetilde G-}^{}}{8c_W^2}, \quad
\tilde\chi_{13}^\ga =-\frac{\eta_{\widetilde G+}^{}}{8c_W^2}
-\frac{\eta_{\widetilde G-}^{}}{8c_W^2},
\nn\\
&\tilde\chi_{30}^\ga =\frac{\eta_{{B}W}^{}}4-\frac{t_W^{}\eta_{{W}W}^{}}{8}
-\frac{\eta_{{B}B}^{}}{2t_W^{}},
\quad 
\tilde\chi_{31}^\ga =\frac{\eta_{\widetilde G-}^{}}{8c_W^2}, \quad
\tilde\chi_{33}^\ga =-\frac{\eta_{\widetilde G+}}{8c_W^2}
-\frac{\eta_{\widetilde G-}}{8c_W^2},
\nn
\end{align}
where  $\,\eta_j^{}\!\!=\hsm\!v^2M_Z^2/[\cut_j^4]$,
$t_W^{}\!\!=\!s_W^{}/c_W^{}$, $t_{2W}^{}\!\!=\hsm\!s_{2W}^{}/c_{2W}^{}$, and
$(s_{2W}^{},\hs c_{2W}^{})\!=\!
(\sin\hsm 2\theta_W^{},\hs\cos\hsm 2\theta_W^{})$.\ 
Using the on-shell conditions for the initial/final state fermions,
we can readily prove that the contributions of 
$X_{1V}^{\beta\mu}$ and $\!X_{3V}^{\alpha\beta}$ 
to the physical processes of the present study vanish, 
so they are not needed for this work, but 
we list them here for completeness.

\vspace*{1mm}

Finally, we clarify the difference between our new nTGC form factor formulation
and the conventional nTGC form factor formulation\,\cite{Gounaris:1999kf}.\
Since there is no conventional form factor formulation for the doubly off-shell nTGVs
$Z^*\ga V^*$ (with only $\ga$ on-shell) given in the 
literature\,\cite{Gounaris:1999kf}\cite{Degrande:2013kka}, we compare our new
formulation and the conventional formulation\,\cite{Gounaris:1999kf}   
for the nTGVs $Z\ga V^*$ when both $Z$ and $\ga$ are on-shell.\ 
The conventional form factor parametrization of the CPC and CPV neutral triple gauge vertices
$\widetilde{\Gamma}_{Z\gamma V^*}^{\alpha\beta\mu}$
yield the following formulae\,\cite{Gounaris:1999kf}:
{\small 
\beqs 
\label{eq:ZAV*-old}
\begin{align}
	\label{eq:ZAV*-CPC-old}
	\hspace{-3mm}
	\widetilde{\Gamma}_{Z\gamma V^*}^{\alpha\beta\mu(\text{CPC})} 
	&\!=
	\frac{\,e\hs (q_3^2\hsm -\!M_V^2)\,}{\,M_Z^2\,}\!
	\[h_3^V q_{2\nu}^{}\epsilon^{\alpha\beta\mu\nu}
	\hsm +\hsm \dis\frac{h_4^V}{\,M_Z^2\,}\hs
	q_2^{\alpha}\hs q_{3\nu}^{}\hs  q_{2\sigma}^{}\hs\epsilon^{\beta\mu\nu\sigma}\]\! ,
	\\[0.4mm]
	\label{eq:ZAV*-CPV-old}
	\widetilde{\Gamma}_{Z\gamma V^*}^{\alpha\beta\mu (\text{CPV})}
	&\!=
	\frac{\,e\hs (q_3^2\hsm -\! M_V^2)\,}{\,M_Z^{2}~}\!
	\[ h_1^V(q_2^\alpha g^{\mu\beta}\hsm\!-\hsm q_2^\mu g^{\alpha\beta})
	+\hsm \frac{h_2^V}{\,2M_Z^2\,}\hs q_2^{\alpha}g^{\mu\beta}
	(M_Z^2-q_3^2)\hsm\]\! .
\end{align}
\eeqs 
We found in~\cite{Ellis:2022zdw} that the conventional CPC form factor $h_4^V$ of 
Eq.\eqref{eq:ZAV*-CPC-old} is incompatible with 
the spontaneous breaking of the full electroweak gauge group
SU(2)$_W^{}\otimes$U(1)$_Y^{}\!\!\to\hs$U(1)$_{\text{em}}$, 
and leads to unphysically large 
high-energy behaviors.\ Hence we must introduce a new form factor $h_5^{V}$ 
as in Eq.\eqref{eq:ZAV*-CPC-h4h5} for matching consistently with
the broken phase results originating from the CPC dimension-8 operators
\eqref{eq:obw-CPC}-\eqref{eq:obwT-CPC} and \eqref{eq:OG+}-\eqref{eq:OG-}  
of the SMEFT that respect the underlying electroweak SU(2)$_W^{}\otimes$U(1)$_Y^{}$ gauge symmetry.\ 
Furthermore, we find that the conventional CPV form factor $h_2^V$ of 
Eq.\eqref{eq:ZAV*-CPV-old} is incompatible with 
the spontaneous breaking of the full electroweak gauge group 
SU(2)$_W^{}\otimes$U(1)$_Y^{}\!\to\hs$U(1)$_{\text{em}}$ 
and also causes unphysically large 
high-energy behavior.\ To construct a consistent formulation, we introduce 
another new form factor $h_6^{V}$ as in Eq.\eqref{eq:ZAV*-CPV-h2h6} 
to match with
the broken phase results originating from the CPV dimension-8 operators
\eqref{eq:obw-CPV}-\eqref{eq:obb-CPV} and \eqref{eq:vOG+}-\eqref{eq:vOG-}
of the SMEFT that respect the full underlying electroweak gauge symmetry.\ 
We then present the extended nTGC form factor formulations 
for both the CPC case\,\cite{Ellis:2022zdw} and the CPV case:
\beqs 
\begin{align}
	\label{eq:ZAV*-CPC-h4h5} 
	\hspace*{-8mm}
	\Gamma_{Z\gamma V^*}^{\alpha\beta\mu(\text{CPC})}
	&\!=
	\frac{\,e\hs (q_3^2\hsm -\!M_V^2)\,}{\,M_Z^2\,}\!
	\[\!\(\!h_3^V\hsm\!+h_5^V\!\frac{q_3^2}{\,M_Z^2\,}\!\)\!
	q_{2\nu}^{}\epsilon^{\alpha\beta\mu\nu}
	\hsm +\hsm \dis\frac{h_4^V}{\,M_Z^2\,}\hs
	q_2^{\alpha}\hs q_{3\nu}^{}\hs  q_{2\sigma}^{}\hs\epsilon^{\beta\mu\nu\sigma}\]\! ,
	\\[1mm]
	\label{eq:ZAV*-CPV-h2h6}
	\hspace*{-8mm}
	\Gamma_{Z\gamma V^*}^{\alpha\beta\mu(\text{CPV})}
	&\!=
	\frac{\,e\hs (q_3^2\hsm -\! M_V^2)\,}{\,M_Z^{2}~}\!
	\[\!\(\! h_1^V\!\hsm + h_6^V\!\frac{\,q_3^2\,}{M_Z^2}\!\)\!\!
	(q_2^\alpha g^{\mu\beta}\hsm\!-\hsm q_2^\mu g^{\alpha\beta})
	\hsm +\! \frac{h_2^V}{\,2M_Z^2\,}\hs q_2^{\alpha}g^{\mu\beta}
	(M_Z^2\!-\!q_3^2)\hsm\]
	\nn\\[-5mm]
	\\[-1mm]
	&\!=
	\frac{\,e\hs (q_3^2\hsm -\! M_V^2)\,}{\,M_Z^{2}\,}\!\hsm
	\[\hsm h_1^V\hsmx (q_2^\alpha g^{\mu\beta}\hsm\!-\hsm 
	q_2^\mu g^{\alpha\beta})\hsmx +\hsmx
	\frac{\,h_2^V\hsm M_Z^2\hs q_2^{\alpha}g^{\mu\beta}
		\!-\!2h_6^Vq_3^2q_2^\mu g^{\alpha\beta}\,}{\,2M_Z^2\,}
	\hsmx +\hsmx
	\frac{\,2h_6^V\hsm\!-\!h_2^V\,}{2M_Z^2}q_3^2q_2^\alpha g^{\mu\beta}\]
	\hsm\! ,
	\nn
\end{align}
\eeqs 
where the form factor coefficients $h_{1,2,3,4}^V$ are 
the conventional form factors\,\cite{Gounaris:1999kf} and the 
form factor terms $\propto h_{5}^V$ and $h_{6}^V$ belong to our extension.\  
We observe that the form factor coefficients $(h_{4}^V,\,h_5^V)$ (CPC)
and $(h_{2}^V,\,h_6^V)$ (CPV) are subject to further constraints due to the matching with the broken phase results of the corresponding dimension-8 nTGC operators
that respect the full SU(2)$_W^{}\otimes$U(1)$_Y^{}$ electroweak gauge symmetry.\ 
These constraints ensure that for each scattering amplitude the nTGC form factor contributions have the same high-energy behaviors as those of the
corresponding dimension-8 nTGC operators.\
For this, we make a precise matching between the above nTGC form factor formulae
\eqref{eq:ZAV*-CPC-h4h5}-\eqref{eq:ZAV*-CPV-h2h6} and the 
dimension-8 nTGC operators \eqref{eq:dim8H}-\eqref{eq:OG+G-}.\  
From these we derive the following nontrivial relations 
between the nTGC form factor coefficients:
\beqs 
\label{eq:cond-h45-h26}
\begin{alignat}{3}
	\label{eq:cond1-h45-h26}
	& h_4^V=2h_5^V \, ,~~&&~~ h_2^V=2h_6^V \, ,
	\\
	\label{eq:cond2-h4-h2}
	& h_4^{Z} = \frac{\,c_W^{}}{\,s_W^{}} h_4^{\ga} \, ,
	~~&&~~ h_2^{Z} = \frac{\,c_W^{}}{\,s_W^{}} h_2^{\ga} \, .
\end{alignat}
\eeqs
Imposing these relations, we can further express the formulae 
\eqref{eq:ZAV*-CPC-h4h5} and \eqref{eq:ZAV*-CPV-h2h6}
as follows:
\beqs 
\label{eq:ZAV*-CPC-h4h5-h1h2}  
\begin{align}
	\label{eq:ZAV*-CPC-h4h5new}  
	\hspace*{-5mm}
	\Gamma_{Z\gamma V^*}^{\alpha\beta\mu(\text{CPC})}
	&\!=
	\frac{\,e\hs (q_3^2\hsm -\!M_V^2)\,}{\,M_Z^2\,}\!
	\[\!\(\!h_3^V\hsm\!+h_4^V\!\frac{q_3^2}{\,2M_Z^2\,}\!\)\!
	q_{2\nu}^{}\epsilon^{\alpha\beta\mu\nu}
	\hsm +\hsm \dis\frac{h_4^V}{\,M_Z^2\,}\hs
	q_2^{\alpha}\hs q_{3\nu}^{}\hs  q_{2\sigma}^{}\hs\epsilon^{\beta\mu\nu\sigma}\]\! ,
	\\[1mm]
	\label{eq:ZAV*-CPV-h2h6new}
	\hspace*{-5mm}
	\Gamma_{Z\gamma V^*}^{\alpha\beta\mu(\text{CPV})} 
	&\!= \frac{\,e\hs (q_3^2\hsm -\! M_V^2)\,}{\,M_Z^{2}\,}\!\hsm
	\[\hsm h_1^V\hsmx (q_2^\alpha g^{\mu\beta}\hsm\!-\hsm 
	q_2^\mu g^{\alpha\beta})\hsmx +
	h_2^V\hsm \frac{~M_Z^2\hs q_2^{\alpha}g^{\mu\beta}
		\!-\!q_3^2q_2^\mu g^{\alpha\beta}\,}{\,2M_Z^2\,}\]
	\! .
\end{align}
\eeqs 

Applying naive power counting on the individual contributions of the nTGC form factors 
of Eqs.\eqref{eq:ZAV*-CPC-h4h5}-\eqref{eq:ZAV*-CPV-h2h6} to the scattering amplitude for
$f\hs\bar{f}\ito Z\ga\hs$, 
we deduce the size of each individual contribution as follows:
\\[-4.5mm]
\beqs
\label{eq:count-h345-h126} 
\begin{align}
	\label{eq:count-CPC-ZT} 
	\TT[Z_T^{}\ga_T^{}](\text{CPC}) &=\,
	h_3^V O(E^2) + h_5^V O(E^4) \,,
	\\
	\label{eq:count-CPC-ZL} 
	\TT[Z_L^{}\ga_T^{}](\text{CPC}) &=\,	
	h_3^V O(E^3) +h_4^V O(E^5)+ h_5^V O(E^5) \,,
	\\[0.5mm]
	\label{eq:count-CPV-ZT} 
	\TT[Z_T^{}\ga_T^{}](\text{CPV}) &=\,
	h_1^V O(E^2) + h_6^V O(E^4) \,,
	\\
	\label{eq:count-CPV-ZL} 
	\TT[Z_L^{}\ga_T^{}](\text{CPV}) &=\,	
	h_1^V O(E^3) +h_2^V O(E^5)+ h_6^V O(E^5) \,.
\end{align}
\eeqs 
We note that the form factors $h_4^V$ and $h_2^V$ do not contribute
to the amplitudes with final state $Z_T^{}\ga_T^{}\,$.\ This is because
the $h_4^V$ and $h_2^V$ vertices in Eqs.\eqref{eq:ZAV*-CPC-h4h5}-\eqref{eq:ZAV*-CPV-h2h6}
contain the momentum 
$q_2^{\alpha}\!=\!-(q_1^{\al}\!+\hsm q_3^{\al})\hs$, and thus 
the contraction $q_1^{\al}\ep_{T\al}^{Z}(q_1^{})$ vanishes due to the
on-shell condition of $Z$ boson, and the other contraction 
$q_3^{\al}\ep_{T\al}^{Z}(q_1^{})$ vanishes due to the fact that 
the $s$-channel momentum $q_3^{\al}$ has no spatial component
and the transverse polarization vector $\ep_{T\al}^{Z}$ has no time component.\

\vspace*{1mm}

Inspecting the scattering amplitudes of 
Eq.\eqref{eq:count-h345-h126}, we note that
for the final state $Z_L^{}\ga_T^{}$, 
the largest individual contributions of $\mO(E^5)$ 
arise from the form factors $(h_4^V,\hs h_5^V)$ for the CPC case, 
as shown in Eq.\eqref{eq:count-CPC-ZL}, and
from the form factors $(h_2^V,\hs h_6^V)$ for the CPV case
as shown in Eq.\eqref{eq:count-CPV-ZL}.\
We have verified explicitly that upon imposing the matching constraints
$h_4^V\!=\!2h_5^V$ (CPC case) 
and 
$h_2^V\!=\!2h_6^V$ (CPV case) of Eq.\eqref{eq:cond1-h45-h26},
the sum of the individual leading contributions 
exhibits an exact energy cancellation $O(E^5)\hsm\ito O(E^3)$
for each amplitude.\ Thus, the nonzero leading contributions behave as
$O(E^4)$ and arise from the final state $Z_T^{}\ga_T^{}$ instead.\ 
The large energy cancellations $O(E^5)\hsm\ito O(E^3)$
can be understood by applying the equivalence theorem (ET)\,\cite{ET}
to the high-energy scattering process $f\hs\bar{f}\ito Z_L^{}\ga_T^{}\hs$.\
For $E\!\gg\! M_Z^{}\hs$, the ET gives:
\beqa
\vspace*{-2mm}
\mathcal{T}[Z_L^{},\ga_T^{}] \,=\,
\mathcal{T}[-\ii\pi^0,\ga_T^{}] + B\,,
\label{eq:ET}
\eeqa
where the longitudinal gauge boson $Z_L^{}$ absorbs
the would-be Goldstone boson $\pi^0$
through the Higgs mechanism, and
the residual term
$\,B\hsm =\hsm\mathcal{T}[v^\mu Z_\mu^{},\ga_T^{}]\,$
is suppressed by the quantity
$\,v^\mu\!\hs\equiv\!\epsilon_L^\mu -$
$q_Z^\mu/M_Z^{}\hsm =\hsm O(M_Z^{}/E_Z^{})$ \cite{ET}.\
We note that the ET \eqref{eq:ET} cannot be directly applied to the
conventional form factor formulation \eqref{eq:ZAV*-old}, because it
obeys only the gauge symmetry U(1)$_{\text{em}}^{}$ and 
does not respect the full electroweak gauge symmetry of the SM.\ 
More generally, the form factor formulation is normally given in the
broken phase of the electroweak gauge symmetry and 
contains no would-be Goldstone boson.\ 
We should apply the ET to the electroweak gauge-invariant formulation
of the nTGCs which can be derived only from the dimension-8 operators
as in Eqs.\eqref{eq:dim8H}-\eqref{eq:OG+G-}.\ 
Thus, we can analyze the allowed
leading energy-dependences of the amplitudes
\eqref{eq:count-CPC-ZL} and \eqref{eq:count-CPV-ZL}
by applying the ET to the contributions
of the dimension-8 nTGC operators in Eqs.
\eqref{eq:dim8H}-\eqref{eq:OG+G-}.\ 
From these, we find that only the Higgs-related nTGC operators
in Eq.\eqref{eq:dim8H} could 
contribute to the Goldstone amplitude
$\,\mathcal{T}[-\ii\pi^0,\ga_T^{}]\,$ 
with a leading energy-dependence $O(E^3)$
that corresponds to the form factor $h_3^Z\hs$ or $h_1^Z\hs$.\
The operators $\hs\mathcal{O}_{G_\pm}^{}\!$
or $\hs\widetilde{\mathcal{O}}_{G_\pm}^{}\!$ 
do not contribute to the Goldstone boson amplitude
$\mathcal{T}[-\ii\pi^0,\ga_T^{}]$,
but they contribute the largest residual term $B\! =\! O(E^3)\hs$.\
With these results, we apply the ET \eqref{eq:ET} and deduce that in 
Eqs.\eqref{eq:count-CPC-ZL} and \eqref{eq:count-CPV-ZL} 
the individual leading terms of $O(E^5)$ due to the form factors
$(h_4^V,\hs h_5^V)$ (CPC case) or
$(h_2^V,\hs h_6^V)$ (CPV case)
must exactly cancel with each other for the on-shell amplitude
$\mathcal{T}[Z_L^{},\ga_T^{}]$,
from which we derive the the following conditions:
\beq
\label{eq:E5cancel-h4h5}
h_4^V\!/\hs h_5^V =\, 2\,,~~~~~~
h_2^V\!/\hs h_6^V =\, 2\,,
\eeq
in agreement with Eq.\eqref{eq:cond1-h45-h26}.\
From the above, we further conclude that the final nonzero leading amplitude 
is $O(E^4)$ as in $\TT[Z_T^{}\ga_T^{}]$ and is
contributed by $h_5^V$ for the CPC case shown in Eq.\eqref{eq:count-CPC-ZT} 
or by $h_6^V$ for the CPV case shown in Eq.\eqref{eq:count-CPV-ZT}.  

\vspace*{1mm}

Next we consider the scattering process 
$f\bar{f}\ito Z^*\ga\ito f'\!\bar{f}'\ga$ 
with an off-shell $Z^*$, as studied in the main text.\
Using our doubly off-shell nTGV formulase in Eqs.(5) and (7),
we can count the leading energy-dependence for the contribution of
each form factor as follows:
\beqs 
\begin{align}
	\label{eq:count-CPC-Z*A} 
	\TT[f'\!\bar{f}'\ga_T^{}](\text{CPC}) &=\,	
	h_{31}^{\ga}O(E^1) +\hh_3^{\ga}O(E^3) + \hh_3^{Z}O(E^1)+
	\hh_4^{}O(E^3) \,,	
	\\[0.5mm]
	\TT[f'\!\bar{f}'\ga_T^{}](\text{CPV}) &=\,	
	h_{11}^{\ga}O(E^1) +\hh_1^{\ga}O(E^3) + \hh_1^{Z}O(E^1)+
	\hh_2^{}O(E^3) \,.
\end{align}
\eeqs 
We see that the nonzero leading energy dependence is $O(E^3)$,
contributed by the form factors $(\hh_3^{\ga},\hs\hh_4^{})$ and 
$(\hh_1^{\ga},\hs\hh_2^{})$.\ 
This is in contrast with the on-shell amplitudes
$\TT[Z_T^{}\ga_T^{}]$ in Eqs.\eqref{eq:count-CPC-ZT} and \eqref{eq:count-CPV-ZT},
which have nonzero leading energy-dependences of $O(E^4)$, 
contributed by $h_5^V$ and $h_6^V$.\
\\[-7mm]

\hspace*{0.5mm}

As shown in Eqs.(5) and (7) 
of Section\,II, 
we find that the contributions of $\hh_{3}^{\ga}$ and $\hh_{1}^{\ga}$ 
to the nTGVs are enhanced by a large factor  
$q_1^2/M_Z^2$ from the off-shell decays 
$Z^*(q_1^{})\!\ito\nu\hs\bar\nu\ga\hs$, whereas the contributions of
$\hh_{3,1}^Z$ and  $\hh_{4,2}^{}$ are dominated by on-shell $Z$ decays,
so including off-shell $Z^*$ decays 
does not cause significant changes in their sensitivity reaches.\  
Hence, for studying the form factors  $\hh_{4}^{}$ and $\hh_{2}^{}\hs$,
we can consider the $Z\ga V^*$ nTGVs, and make a quantitative comparison 
between our new form factor formulation 
\eqref{eq:ZAV*-CPC-h4h5-h1h2} and the conventional form factor formulae 
\eqref{eq:ZAV*-old} \cite{Gounaris:1999kf} via the production channel 
$pp(q\bar{q})\ito Z\ga\ito \nu\hs\bar\nu\ga$\, at hadron colliders.\ 
We present this comparison in
Table\,\ref{tab:3}, where the third row (red color) depicts the sensitivity reaches 
on $\hh_{4,2}^{}$ obtained by using our nTGV formula \eqref{eq:ZAV*-CPC-h4h5-h1h2},
and the last two rows (blue color) show the sensitivity reaches 
on $h_{4,2}^{V}$  that would be obtained
by using the (incorrect) conventional nTGV formulas \eqref{eq:ZAV*-old}~\cite{Gounaris:1999kf}.\
Table\,\ref{tab:3} demonstrates that the sensitivity reaches on 
the conventional nTGC form factors $h_{4,2}^{V}$ 
are erroneously stronger than those for our
new formulation \eqref{eq:ZAV*-CPC-h4h5-h1h2} by about a factor of $O(20)$
at the LHC and by about a factor of $O(170)$ at a 100{\hs}TeV $pp$ collider.\ 

\vspace*{1mm}  

Finally, we comment on the possible variations of the operator structure
$F^2H^2D^2$.\ 
We note that there are two subclasses of $F^2H^2D^2$ 
which can contribute to nTGCs\,\cite{Degrande:2013kka}: 
$H^\dagger F_{\mu\nu}F^{\mu\rho}D_\rho D^\nu H$ and 
$H^\dagger D_\rho F_{\mu\nu}F^{\mu\rho} D^\nu H$.\ 
Using integration by parts, we can convert the operator 
$H^\dagger D_\rho F_{\mu\nu}F^{\mu\rho} D^\nu H$  to 
$H^\dagger F_{\mu\nu}F^{\mu\rho}D_\rho D^\nu H$ plus certain other operators.\ 
The operators of the subclass $H^\dagger F_{\mu\nu}F^{\mu\rho}D_\rho D^\nu H$ 
were shown in \cite{Degrande:2013kka}.\ 
We further present the subclass 
$H^\dagger D_\rho F_{\mu\nu}F^{\mu\rho} D^\nu H$ 
which includes the following operators:  
\begin{align}
	\label{eq:OFF}
	\mO_{BW}^{(1)}& =\ii  H^\dagger D_\rho \widetilde B_{\mu\nu}W^{\mu\rho} D^\nu H+\text{h.c.},\quad
		\mO_{WB}^{(1)} =\ii H^\dagger D_\rho \widetilde  W_{\mu\nu} B^{\mu\rho} D^\nu H+\text{h.c.},
	\nn\\ 
	\mO_{BB}^{(1)} & =\ii H^\dagger D_\rho\widetilde  B_{\mu\nu}B^{\mu\rho} D^\nu H+\text{h.c.},\quad \mO_{WW}^{(1)}=iH^\dagger D_\rho \widetilde W_{\mu\nu} W^{\mu\rho} D^\nu H+\text{h.c.},
	\nn\\
	\mO_{BB}^{(2)} & =\ii H^\dagger D_\rho B_{\mu\nu}\widetilde B^{\mu\rho} D^\nu H+\text{h.c.},\quad 
	\mO_{WW}^{(2)} =\ii H^\dagger D_\rho W_{\mu\nu}\widetilde W^{\mu\rho} D^\nu H+\text{h.c.},
	\\
	\widetilde{\mO}_{BW}^{(1)} &=\ii H^\dagger D_\rho B_{\mu\nu}W^{\mu\rho} D^\nu H+\text{h.c.},\quad 
	\widetilde{\mO}_{WB}^{(1)} =\ii H^\dagger D_\rho W_{\mu\nu}B^{\mu\rho} D^\nu H+\text{h.c.},
	\nn\\ 
	\widetilde{\mO}_{BB}^{(1)} &=\ii H^\dagger D_\rho B_{\mu\nu}B^{\mu\rho} D^\nu H+\text{h.c.},\quad
	\widetilde{\mO}_{WW}^{(1)} =\ii H^\dagger D_\rho W_{\mu\nu} W^{\mu\rho} D^\nu H+\text{h.c.}.
	\nn 
\end{align}
We note that in addition to the above operators, there are two other operators
$\mO_{BW}^{(2)}\!=\!
\ii  H^\dagger D_\rho B_{\mu\nu}\widetilde W^{\mu\rho} D^\nu H$
and
$\mO_{WB}^{(2)}\!=\!\ii H^\dagger D_\rho W_{\mu\nu}\widetilde B^{\mu\rho} D^\nu H$
belonging to the same subclass of $H^\dagger D_\rho F_{\mu\nu}F^{\mu\rho} D^\nu H$.\  
But, by using the Bianchi identity $D_\rho\widetilde F^{\mu\rho}\!=\!0$\, 
and integration by parts, we find that they contribute the same nTGVs 
as that of $\mO_{\widetilde{B}W}^{}$ in Eq.(2.2a).\ 
So they are not independent and thus not included in Eq.\eqref{eq:OFF}.\
We can expand the above operators \eqref{eq:OFF} and derive their contributions
to the nTGVs:
\beqs
\label{eq:O-II}
\beqa
{\mO}_{VV}^{(n)}&=&-\frac{v^2 e Z_\nu}{2s_W c_W}(a_1\partial_{\rho}\widetilde A_{\mu\nu}A^{\mu\rho}+a_2\partial_{\rho}\widetilde Z_{\mu\nu}Z^{\mu\rho}+a_3\partial_{\rho} A_{\mu\nu}\widetilde A^{\mu\rho}+a_4\partial_{\rho} Z_{\mu\nu}\widetilde Z^{\mu\rho}\nn\\
&&+a_5\partial_{\rho}\widetilde A_{\mu\nu}Z^{\mu\rho}+a_6\partial_{\rho}\widetilde Z_{\mu\nu}A^{\mu\rho}+a_7\partial_{\rho} A_{\mu\nu}\widetilde Z^{\mu\rho}
+a_8\partial_{\rho} Z_{\mu\nu}\widetilde A^{\mu\rho})\hs ,
\\[1mm] 
\widetilde{\mO}_{VV}^{(n)}&=&-\frac{v^2 e Z_\nu}{2s_W c_W}(b_1\partial_{\rho} A_{\mu\nu}A^{\mu\rho}+b_2\partial_{\rho}Z_{\mu\nu}Z^{\mu\rho}+b_3\partial_{\rho} A_{\mu\nu}Z^{\mu\rho}+b_4\partial_{\rho}Z_{\mu\nu}A^{\mu\rho}) \hs,
\eeqa
\eeqs
where the vertices ${\mO}_{VV}^{(n)}$ are CPC and $\widetilde{\mO}_{VV}^{(n)}$
are CPV, with the index $n\!=\!1,2\hs$.\ 
In the above, the coefficients $(a_j^{},\,b_j^{})$ are given by 
\pagebreak

\beqs
\vspace*{-1cm}
\beqa
&&a_1=-\frac{c^{(1)}_{BW}c_W s_W}{2}-\frac{c^{(1)}_{WB}c_W s_W}{2}+c^{(1)}_{BB}c_W^2+\frac{c^{(1)}_{WW} s_W^2}{4},\\
&&a_2=\frac{c^{(1)}_{BW}c_W s_W}{2}+\frac{c^{(1)}_{WB}c_W s_W}{2}+c^{(1)}_{BB}s_W^2+\frac{c^{(1)}_{WW} c_W^2}{4},\\
&&a_3=c^{(2)}_{BB}c_W^2+\frac{c^{(2)}_{WW} s_W^2}{4},\\
&&a_4=c^{(2)}_{BB}s_W^2+\frac{c^{(2)}_{WW} c_W^2}{4},\\
&&a_5=-\frac{c^{(1)}_{BW}c_W^2}{2}+\frac{c^{(1)}_{WB}s_W^2}{2}-c^{(1)}_{BB}c_Ws_W+\frac{c^{(1)}_{WW} c_Ws_W}{4},\\
&&a_6=\frac{c^{(1)}_{BW}c_W^2}{2}-\frac{c^{(1)}_{WB}s_W^2}{2}-c^{(1)}_{BB}c_Ws_W+\frac{c^{(1)}_{WW} c_Ws_W}{4},\\
&&a_7=-c^{(2)}_{BB}c_Ws_W+\frac{c^{(2)}_{WW} c_Ws_W}{4},\\
&&a_8=-c^{(2)}_{BB}c_Ws_W+\frac{c^{(2)}_{WW} c_Ws_W}{4},
\eeqa 
\eeqs 
and
\beqs 
\beqa 
&&b_1=-\frac{\tilde c^{(1)}_{BW}c_W s_W}{2}-\frac{\tilde c^{(1)}_{WB}c_W s_W}{2}+\tilde c^{(1)}_{BB}c_W^2+\frac{\tilde c^{(1)}_{WW} s_W^2}{4},\\
&&b_2=\frac{\tilde c^{(1)}_{BW}c_W s_W}{2}+\frac{\tilde c^{(1)}_{WB}c_W s_W}{2}+\tilde c^{(1)}_{BB}s_W^2+\frac{\tilde c^{(1)}_{WW} c_W^2}{4},\\
&&b_3=-\frac{\tilde c^{(1)}_{BW}c_W^2}{2}+\frac{\tilde c^{(1)}_{WB}s_W^2}{2}-\tilde c^{(1)}_{BB}c_Ws_W+\frac{\tilde c^{(1)}_{WW} c_Ws_W}{4},\\
&&b_4=\frac{\tilde c^{(1)}_{BW}c_W^2}{2}-\frac{\tilde c^{(1)}_{WB}s_W^2}{2}-\tilde c^{(1)}_{BB}c_Ws_W+\frac{\tilde c^{(1)}_{WW} c_Ws_W}{4}.
\eeqa
\eeqs
Thus, we derive their contribution to the nTGC form factors:
\beqs 
\begin{align}
	\hat h_3^Z & =\frac{v^2M_Z^2}{\,2c_W^{}s_W^{}\cut^4\,}
	(-a_5^{}-a_6^{}-a_7^{}+a_8^{})\,,
	\\
	h_{31}^\ga &=-\frac{v^2M_Z^2}{\,c_W^{}s_W^{}\cut^4\,}a_1^{} \,,
	\\
	\hat h_1^Z &=\frac{v^2M_Z^2}{\,4c_W^{}s_W^{}\cut^4\,}(b_4^{}-b_3^{}) \,.
\end{align}
\eeqs 
Inspecting Eq.\eqref{eq:O-II}, we find that the subclass $H^\dagger D_\rho F_{\mu\nu}F^{\mu\rho} D^\nu H$ 
contribute the same Lorentz structures of the form factors as that
of the other subclass $H^\dagger F_{\mu\nu}F^{\mu\rho}D_\rho D^\nu H$.\
Hence it will suffice to focus our analysis just on one kind of subclass operators 
$H^\dagger F_{\mu\nu}F^{\mu\rho}D_\rho D^\nu H$
in the main text.
\begin{table}[t]
	\begin{center}
		\begin{tabular}{c|ccc||c|ccc}
			\hline\hline	
			& & & & & & &
			\\[-4mm]
			$\sqrt{s\,}$ & & \hspace*{-12mm}13\,TeV\,
			\hspace*{-12mm}
			& & & & &\hspace*{-16mm}100\,TeV\
			\\
			\hline
			& & & & & & &
			\\[-4.3mm]
			$\mL$(ab$^{-1}$) & 0.14 & 0.3 & 3 & 
			& 3 & 10 & 30 			
			\\
			\hline\hline
			& & & & & & &
			\\[-4.3mm]
			\red$|\hat h_{4,2}^{}|\!\times\!10^{6}$
			& \red 11\, & \red 8.5\, & \red 4.2\, &
			\red $|\hat h_{4,2}^{}|\!\times\!10^{9}$
			& \red 4.5 &\red 2.9 & \red 2.0
			\\
			& & & & & & &
			\\[-4.3mm]
			\hline
			& & & & & & &
			\\[-4.3mm]
			\blu $| h_{4,2}^Z|\!\times\!10^{6}$
			&\blu 0.47 &\blu  0.37 &\blu  0.19
			&\blu   $| h_{4,2}^Z|\!\times\!10^{11}$
			&\blu  2.6 &\blu  1.7 &\blu  1.2
			\\
			& & & & & & &
			\\[-4.3mm]
			\hline
			& & & & & & &
			\\[-4.3mm]
			$\blu | h_{4,2}^\ga|\!\times\!10^{6}$
			&\blu 0.54 &\blu 0.43 &\blu 0.22
			& $\blu | h_{4.2}^\ga|\!\times\!10^{11}$
			&\blu 2.9 &\blu 1.9 &\blu 1.4
			\\[-4.3mm]
			& & & & & & &
			\\
			\hline\hline
		\end{tabular}
	\end{center}
	\vspace*{-4.5mm}
	\caption{\small\hspace*{-2.5mm}
		{\it Comparisons of the $2\hs\sigma$ sensitivity reaches 
			for the form factors $(\hh^{}_4,\hs\hh_2^{})$ formulated in 
			the SMEFT (in red color) and for the conventional form 
			factors $(h^{V}_4,\hs h_2^{V})$ respecting only U(1)$_{\rm{em}}^{}$  (in blue color), derived from the reaction 
			$\,p{\hs}p{\hs}(q{\hs}\bar{q})\hsm\ito Z\ga\hsm\ito\nu\bar{\nu}\ga$		
			at the} LHC\,(13\,TeV) {\it and a} 100\,TeV $pp$ 
		{\it collider, with the indicated integrated luminosities.\ 
			As discussed in the text, the form-factor limits in blue color 
			are included for illustration only,
			as they are incompatible with the spontaneous breaking of the
			SM electroweak gauge symmetry, and hence are invalid.
	}}
	\label{tab:3}
\end{table}

\section{\large\hspace{-6.5mm}.\hspace*{1.5mm}Cross Sections for CPC and CPV nTGC Contributions}
\label{app:sigma-012}
\label{app:B}
\setcounter{equation}{23}

In this section, we present the cross section formulae for
the reaction $\,q\bar q\hsm\ito\hsm Z^*\ga$ in terms of the
SM contribution ($\sigma_0^{}$), the interference between the SM and 
the nTGC contributions ($\sigma_1^{}$), and the squared part of 
the nTGC contributions ($\sigma_2^{}$),
which are used in Eq.(10)
of the main text:
%
\begin{align}
	\label{eq:CSsum-qq-Zgamma}
	\sigma(q\bar q\hsm\ito\hsm Z^*\ga ) \,=&~
	\sigma_0^{} +\sigma_1^{} +\sigma_2^{}\,.
\end{align}
The SM cross section term ($\sigma_0^{}$) and the interference term ($\sigma_1^{}$)
are given by
\beqs
\begin{align}
	\label{eq:CS-qqZA-d8other}
	\sigma_0^{} \,=&\,
	\frac{~e^4(q_L^2\!+\!q_R^2)Q^2\!
		\left[-(\shat\!-\!M_{Z^*}^2)^2\!-\!2(\shat^2\!+\!M_{Z^*}^4)
		\ln\sin\!\frac{\delta }{2}\,\right]\,}
	{\,8\pi s_W^2c_W^2(\shat\!-\!M^2_{Z^*})\hs \shat^2\,}\,,
	\hspace*{15mm}
	\\[1.5mm]
	\label{eq:CS1-qq-Zgamma-H34}
	\sigma_{1}^{} 
	\,=&\,
	-\frac{\,e^4Qq_L^{}\hs{T_3^{}}\!
		\left(\shat\!-\!M_{Z^*}^2\right)\,}{16\pi s_W^2c_W^2M_{Z^*}^2\,\shat} \bar h_{4}^{}
	-\frac{\,e^4 Q(q_L^{}x_L^Z\!\!-\!q_R^{}x_R^Z)
		(\shat^2\!\hsm-\!M_{Z^*}^4)\,}
	{\,16\pi s_W^2c_W^2 M_{Z^*}^2\hs \shat^2\,}\bar h_{3}^Z
	\nn\\[1mm]
	& \,+
	\frac{\,e^4Q(q_L^{}x_L^A\!\!-\!q_R^{}x_R^A)
		(\shat^2\!\hsm-\!M_{Z^*}^4)\,}
	{\,16\pi s_W^3c_W M_{Z^*}^2\hs\shat^2\,}\bar h_{3}^\gamma \,,
\end{align}
\eeqs
while the squared contribution ($\sigma_2^{}$) includes the following terms:
\beqs
\label{eq:CS2-qq-Zgamma-H34}
\begin{align}
	\label{eq:sigma2=sum}
	\sigma_{2}^{} \,=&\,~
	\sigma_{2}^{44}+\sigma_{2Z}^{33} + \sigma_{2A}^{33}
	+\sigma_{2Z}^{43}+\sigma_{2A}^{43}+\sigma_{2Z\hsm A}^{33}\,,
	\\[1.5mm]
	\label{eq:sigma2-44}
	\sigma_{2}^{44}
	\,=&\,
	\frac{~e^4 T_3^2 (\shat\hsm +\!M_{Z^*}^2)(\shat\hsm -\!M_{Z^*}^2){}^3~}
	{~768\hs\pi s_W^2c_W^2M_{Z^*}^8\,\shat~}
	\big[\hsm(\bar{h}_{2}^{}\hsm )^2\!+\!(\bar{h}_{4}^{}\hsm )^2\big] \, ,
	\\[1mm]
	\label{eq:sigma2Z-33}
	\sigma_{2Z}^{33}
	\,=&\,
	\frac{~e^4[Q^2 s_W^4\!+\!(T_3^{}\!-\hsm Q s_W^2\hsm )^{2}]
		(\shat\!+\!M_{Z^*}^2)(\shat\!-\!M_{Z^*}^2)^3\,}
	{~192\hs\pi\hs s_W^2c_W^2M_{Z^*}^6\, \shat^2~}
	\big[\hsm(\bar{h}_{1}^Z\hsm )^2\!+\!(\bar{h}_{3}^Z\hsm )^2\big] \, ,
	\\[1mm]
%
%
	\label{eq:sigma2A-33}
	\sigma_{2A}^{33} \,=&\,
	\frac{~e^4Q^2(\shat\!+\!M_{Z^*}^2)(\shat\!-\!M_{Z^*}^2)^3\,}
	{96\hs\pi M_{Z^*}^6\,\shat^2}
	\big[\hsm(\bar{h}_{1}^{\ga}\hsm )^2\!+\!(\bar{h}_{3}^{\ga}\hsm )^2\big] \, ,
	\\[1mm]
	\label{eq:sigma2Z-43}
	\sigma_{2Z}^{43}
	\,=&\,
	\frac{~e^4 T_3^{}(T_3^{}\!-\!Qs_W^2)(\shat\!-\!M_{Z^*}^2)^3\,}
	{96\hs\pi\hs s_W^2c_W^2 M_{Z^*}^6\,\shat}
	\big(\bar{h}_2^{}\bar{h}_{1}^Z\!+\!\bar{h}_4^{}\bar{h}_{3}^Z\big) \, ,
	\\[1mm]
%
	\label{eq:sigma2A-43}
	\sigma_{2A}^{43}	\,=&\,
	\frac{~e^4Q\hs T_3^{}\hs (\shat\!-\!M_{Z^*}^2)^3\,}
	{~96\hs\pi\hs s_W^{}c_W^{}M_{Z^*}^6\,\shat~}
	\big(\bar{h}_2^{}\bar{h}_{1}^{\ga}\!+\!\bar{h}_4^{}\bar{h}_{3}^{\ga}\big) \, ,
	\\[1mm]
	\label{eq:sigma2ZA-43}
	\sigma_{2ZA}^{33}  \,=&\,
	\frac{~e^4 Q(T_3^{}\!-\!2Qs_W^2)
		(\shat\!+\!M_{Z^*}^2)(\shat\!-\!M_{Z^*}^2)^3\,}
	{~96\hs\pi\hs s_W^{}c_W^{}M_{Z^*}^6 \hs\shat^2~}
	\big(\bar{h}_1^Z\bar{h}_{1}^{\ga}\!+\!\bar{h}_3^Z\bar{h}_{3}^{\ga}\big) \, ,
\end{align}
\eeqs
where the nTGC parameters
\begin{eqnarray}
	\bar h_{1,3}^{Z}=\hat h_{1,3}^Z \frac{M_{Z^*}^2}{M_Z^2}\hs , \qquad
	\bar h_{1,3}^{\ga}= h_{1,3} \frac{M_{Z^*}^2}{M_Z^2}+\hat h_{1,3}\frac{M_{Z^*}^4}{M_Z^4}\hs ,\qquad
	\bar h_{2,4}^{}=\hat h_{2,4} \frac{M_{Z^*}^4}{M_Z^4} \, .
\end{eqnarray}
In the above,  the squared-mass
$M_{Z^*}^2\!\!=\hsm q_{1}^2$ with $q_1^{}$ being the $Z^*$ momentum, 
and the coefficients
$(q_L^{},\,q_R^{})\!=\! (T_3^{}\hsmx -\hsm Qs_W^2,\hs -Qs_W^2)$,
$(x_L^Z,\hs x_R^Z)\!=\!(T_3^{}\hsmx -\hsm Qs_W^2,\hs -Qs_W^2)$, and
$(x_L^A,\hs x_R^A)\!=\!-Qs_W^2(1,\hs 1)$
denote the (left,\,right)-handed gauge couplings between
quarks and the $Z$ boson.\

%
\begin{table}[t]
	\begin{center}
		\begin{tabular}{c|ccc||c|ccc}
			\hline\hline	
			& & & & & &
			\\[-4mm]
			$\sqrt{s\,}$ & & \hspace*{-12mm}13\,TeV\,
			\hspace*{-12mm}
			& & & &
			&\hspace*{-16mm}100\,TeV\
			\\
			\hline
			& & & & & &
			\\[-4.3mm]
			$\mL$(ab$^{-1}$) & 0.14 & 0.3 & 3 & 
			& 3 & 10 & 30 			
			\\
			\hline\hline
			& & & & & &
			\\[-4.3mm]
			$|\hat h_{4,2}^{}|\!\times\!10^{6}$
			& 9.6\, & 7.5\, & 3.8\,& $|\hat h_{4,2}^{}|\!\times\!10^{9}$
			& 3.9 & 2.6 & 1.8
			\\
			\hline
			& & & & & &
			\\[-4.3mm]
			$|\hat h_{3,1}^Z|\!\times\!10^{4} $
			& 1.9 & 1.5 & 0.80&$|\hat h_{3,1}^Z|\!\times\!10^{7} $
			& 6.1 & 4.2 & 3.0
			\\
			\hline
			& & & & & &
			\\[-4.3mm]
			$\red |\hat h_{3,1}^\ga|\!\times\!10^{4}$
			&\red  1.6 &\red  1.2 &\red  0.65
			&\red  $|\hat h_{3,1}^\ga|\!\times\!10^{7}$
			&\red  0.94 &\red  0.62 &\red 0.44\\
			\hline
			& & & & & &
			\\[-4.3mm]
			$\blu | h_{31,11}^\ga|\!\times\!10^{4}$
			&\blu 2.2 &\blu 1.8 &\blu 0.94
			& $\blu | h_{31,11}^\ga|\!\times\!10^{7}$
			&\blu 7.1 &\blu 4.9 &\blu 3.5
			\\
			\hline\hline
		\end{tabular}
	\end{center}
	\vspace*{-5mm}
	\caption{\small\hspace*{-2.5mm}
		{\it Combined sensitivity reaches on probing the CPC and CPV nTGC form factors
			at the $2\hs\sigma$ level, as obtained by analyzing the reactions
			$\,p{\hs}p{\hs}(q{\hs}\bar{q})\hsm\ito\hsm Z^*\ga\hsm\ito\hsm \nu\bar{\nu}\ga$ and $\,p{\hs}p{\hs}(q{\hs}\bar{q})\hsm\ito\hsm Z\ga\hsm\ito\hsm \ell^+\ell^-\ga$		
			at the} LHC\,(13\,TeV) {\it and the} 100\,TeV $pp$
		{\it collider, for the indicated integrated luminosities.\
			In the last two rows, the $\hh_{3,1}^{\ga}$ sensitivities (red color)
			are significantly higher than those of $h_{31,11}^{\ga}$ (blue color)
			because the former contains the off-shell contributions from $\nu\bar{\nu}\ga$ 
			channel as enhanced by $Z^*$-momentum-square ($q_1^2$) and the latter does not.}}
	\label{tab:l+nu}
	\label{tab:4}
\end{table}
%


Finally, we combine the sensitivity reaches on probing the CPC and CPV nTGC form factors
by using the bounds derived from the reaction 
$\,p{\hs}p{\hs}(q{\hs}\bar{q})\hsm\ito\hsm Z^*\ga\hsm\ito\hsm \nu\bar{\nu}\ga$ 
(given in Table\,I of the main text) and the reaction
$\,p{\hs}p{\hs}(q{\hs}\bar{q})\hsm\ito\hsm Z\ga\hsm\ito\hsm \ell^+\ell^-\ga$.\
We analyzed the $\nu\bar{\nu}\ga$ channel in \cite{Ellis:2022zdw} only for CPC nTGCs,
and for the current combined analysis we further include the CPV nTGCs.\
We present the combined sensitivity reaches on probing the CPC and CPV nTGC 
form factors at the $2\hs\sigma$ level,  
for both the on-going LHC and the projected 100\,TeV $pp$ collider,
as shown in Table\,\ref{tab:4}.\
We find that the sensitivity reaches on the form factors $\hh_{4,2}^{}$,
$\hh_{3,1}^Z$, and $h_{31,11}^{\ga}$ are comparable for both
$\nu\bar{\nu}\ga$ and $\ell^+\ell^-\ga$ channels,
where the $\nu\bar{\nu}\ga$ channel has higher sensitivities 
than the $\ell^+\ell^-\ga$ channel by about $(20\!-\!30)$\%
at the LHC and the 100\,TeV $pp$ collider,   
and thus their combined sensitivities have visible improvements.\ 
On the other hand, the sensitivity reaches on the form factors $\hh_{3,1}^{\ga}$
via the $\nu\bar{\nu}\ga$ channel are significantly higher than those via the 
$\ell^+\ell^-\ga$ channel by about 93\% at the LHC and
by a large factor of $\sim\hsm\!11$ at the 100\,TeV $pp$ collider,
because of the major off-shell enhancement on the  $\nu\bar{\nu}\ga$ signals 
as we demonstrated in the main text.\ This explains why the combined sensitivity
reaches on $\hh_{3,1}^{\ga}$ (shown in Table\,\ref{tab:4} and marked in red color)
are dominated by the $\nu\bar{\nu}\ga$ channel and remain nearly the same as 
the sensitivities of the $\nu\bar{\nu}\ga$ channel alone 
(shown in Table\,I of the main text).\ 
We also analyzed the combined sensitivity reaches on the new physics cutoff scale 
$\cut_j^{}$ for each given nTGC operator $\mO_j^{}$.\ Since the nTGC contribution to 
the signal cross section is dominated by the $\sigma_2^{}$ term 
which is proportional to $1/\cut_j^8\hs$ 
(as compared to $\hs\sigma_2^{}\!\propto\!\hh_j^2\hs$ 
for the form factor contributions), 
we would expect a rather minor enhancement by 
$\hs 2^{1/16}\!-\hsm 1\!\simeq\hsm 4.4\%$
even if the two channels of $\,\nu\bar{\nu}\ga\,$ and $\hs\ell^+\ell^-\ga\,$ 
would contribute equal statistic significance.\ 
We present in Table\,\ref{tab:5} the combined sensitivity reaches 
on the new physics scales $\cut_j^{}$} (in TeV) of the dimension-8 nTGC operators
at $2\hs\sigma$ level.\ 
As expected, it shows that the combination of both
$\nu\bar{\nu}\ga$ and $\ell^+\ell^-\ga$ channels leads to fairly minor
improvements on the sensitivity reaches.\  Especially, there are essentially
no visible improvements beyond the sensitivity bounds of the $\nu\bar{\nu}\ga$ channel
(shown in Table\,II of the main text) for new physics scales  
$(\cut_{G-}^{},\,\cut_{\widetilde G-}^{})$, 
which correspond to the nTGC form factors $(\hh_3^{\ga},\,\hh_1^{\ga})$.\ 
This is because the $\nu\bar{\nu}\ga$ channel has large off-shell enhancements
for the contributions of $(\hh_3^{\ga},\,\hh_1^{\ga})$ and thus the
corresponding new physics scale 
$(\cut_{G-}^{},\,\cut_{\widetilde G-}^{})$,
which lead to significantly higher sensitivity reaches than those of the
$\hs\ell^+\ell^-\ga\,$ channel.

\begin{table}[t]
\begin{center}
	\begin{tabular}{c||ccc|ccc}
		\hline\hline	
		&&&&&&
		\\[-3mm]
		$\sqrt{s\,}$ & & \hspace*{-12mm}13\,TeV\,
		\hspace*{-12mm}
		& & & &\hspace*{-17mm}100\,TeV\
		\\
		\hline
		&&&&&&
		\\[-3mm]
		$\mL$\,(ab$^{-1}$) & 0.14 & 0.3 & 3 &  3 & 10 & 30 			
		\\
		\hline\hline
		&&&&&&
		\\[-3mm]
		$\cut_{G+}${\small (CPC)} & 3.4 & 3.6 & 4.2 & 23 & 26 & 29
		\\
		\hline
		&&&&&&
		\\[-3mm]
		$\cut_{G-}^{}${\small (CPC)} & 1.2 & 1.3 & 1.5 & 7.7 & 8.5 & 9.3
		\\
		\hline
		&&&&&&
		\\[-3mm]
		$\cut_{\widetilde{B}W}^{}${\small (CPC)}
		& 1.3 & 1.4 & 1.6 & 5.6 & 6.1 & 6.6
		\\
		\hline
		&&&&&&
		\\[-3mm]
		$\cut_{\widetilde{BW}}${\small (CPC)}
		& 1.5 & 1.6 & 1.9 & 6.4 & 7.0 & 7.6
		\\
		\hline\hline
		&&&&&&
		\\[-3mm]
		$\cut_{\widetilde G+}^{}${\small (CPV)}
		& 2.8 & 3.0 & 3.5 & 20 & 22 &24\\
		\hline
		&&&&&&
		\\[-3mm]
		$\cut_{\widetilde G-}^{}${\small (CPV)}
		& 1.0 & 1.1 & 1.3 & 6.5 & 7.2 & 7.8
		\\
		\hline
		&&&&&&
		\\[-3mm]
		$\cut_{WW}^{}${\small (CPV)}  & 0.96 & 1.0 & 1.2 & 4.0 & 4.4 & 4.8
		\\
		\hline
		&&&&&&
		\\[-3mm]
		$\cut_{WB}^{}${\small (CPV)}
		& 1.1 & 1.2 & 1.4 & 4.8 & 5.2 &5.7\\
		\hline
		&&&&&&
		\\[-3mm]
		$\cut_{BB}^{}${\small (CPV)}
		& 1.4 & 1.5 & 1.7 & 5.8& 64 & 7.0
		\\
		\hline\hline
	\end{tabular}
\end{center}
\vspace*{-5mm}
\caption{\small {\it
		Combined sensitivity reaches on the new physics scales $\cut_j^{}$} (in TeV)
	{\it of the dimension-8 nTGC operators
		at $2\hs\sigma$ level, as obtained by analyzing the reactions
		$\,p{\hs}p{\hs}(q{\hs}\bar{q})\!\ito\! Z^*\ga\!\ito\!\nu\bar{\nu}\ga$  and $\,p{\hs}p{\hs}(q{\hs}\bar{q})\!\ito\! Z\ga\!\ito\hsmx\ell^+\ell^-\ga$		
		at the} LHC\,(13\,TeV) {\it and at a} 100\,TeV $pp$
	{\it collider, with integrated luminosities $\mL$ as indicated.}}
\label{tab:5}
\end{table}

\vspace*{3mm}
\section{\large\hspace{-6.5mm}.\hspace*{1.5mm}Unitarity Constraints on CPC and CPV nTGCs}
\label{app:C}
\setcounter{equation}{27}

In this section, we analyze the perturbative unitarity bounds on
both the CPC and CPV nTGCs through the on-shell scattering process
$\,f\bar{f}\ito Z\ga\,$ with $f\bar{f}=q{\hs}\bar{q}\hs ,\, e^-e^+$.\ 
This also extends our previous unitarity analysis for the CPC nTGCs 
alone\,\cite{Ellis:2022zdw}.\ 
For the scattering amplitude of the reaction $f\bar{f}\ito Z\ga\hs$, 
We make the partial-wave expansion of the nTGC contributions:
%
\begin{eqnarray}
a_J^{}\,=\,\frac{1}{\,32\pi\,}e^{i(\nu'-\nu)\phi}\!\!
\int_{-1}^{1}\!\!\d (\cos\theta)\,
d^{\hs J}_{\nu'\nu}(\cos\theta)\,
\mT_{\text{nTGC}}^{s_f^{}s_{\bar f}^{},\lambda_Z^{}\lambda_\gamma^{}},
\end{eqnarray}
where the differences of initial/final state helicities are given by
$\,\nu \!=s_f^{} -s_{\bar f}^{}\!=\pm 1$ and
$\,\nu'\!=\lambda_Z^{}\!-\!\lambda_\gamma^{}\!=0,\pm 1\hs$,
respectively.\
For the current collider analysis it is sufficient
to treat the initial-state fermions $(f,\bar{f})$
(light quarks or leptons) as massless.\ So we deduce 
$\,s_f^{}\!=\!-s_{\bar f}^{}\,$, leading to
$\,\nu\hsm =\!\pm 1\hs$.\ This means that the
$\hs J\!=\!1\hs$ partial wave gives the leading contribution.\
The relevant Wigner $d$ functions are given by
%
$d^1_{1,0}\!=\!-\fr{1}{\,\sqrt{2\,}\,}\sin\theta\hs$
and $d^1_{1,\pm1}\!=\!\fr{1}{\,2\,}(1\hsm\pm\hsm\cos\theta)\hs$,
%
whereas the general relation
$\,d^{\hs J}_{m,m'}\!=\hsm d^{\hs J}_{-m,-m'}\hs$ holds.\

\begin{table}[t]
\begin{center}
	\begin{tabular}{c||c|c|c|c|c|c|c}
		\hline\hline	
		& & & & &
		\\[-4mm]
		$E_{\rm{CM}}^{}\hs$(TeV) & 0.25&0.5 & 1&3 &5&25&40
		\\[0.3mm]
		\hline\hline
		& & & & & &
		\\[-4.3mm]
		$\Lambda_{G+}^{}$ 
		& 0.078 &0.16 & 0.31 &0.93&1.6&7.8&12 \\
		\hline
		& & & & &&
		\\[-4.3mm]
		$\Lambda_{G-}^{}$ 
		&  0.050 & 0.084 & 0.14 & 0.32 & 0.47 & 1.6&2.2\\
		\hline
		& & & & & &
		\\[-4.3mm]
		$\Lambda_{\!\widetilde{B}W}^{}$  
		&  0.058 & 0.098 & 0.16 & 0.37 & 0.55& 1.8&2.6\\
		\hline
		& & & & &&
		\\[-4.3mm]
		$\Lambda_{\widetilde{BW}}^{}$ 
		&  0.069 & 0.12 & 0.20 & 0.44 & 0.65& 2.2 &3.1\\
		\hline
		& & & & & &
		\\[-4.3mm]
		$\Lambda_{\widetilde G+}^{}$ 
		&  0.065 & 0.13 & 0.26 & 0.79& 1.3 & 6.5&10\\
		\hline
		& & & & &&
		\\[-4.3mm]
		$\Lambda_{\widetilde G-}$ 
		&  0.042 & 0.071 & 0.12 & 0.27 & 0.40 & 1.3&1.9\\
		\hline
		& & & & &&
		\\[-4.3mm]
		$\Lambda_{WW}$ 
		&  0.041 & 0.069 & 0.12 & 0.26 & 0.39 & 1.3&1.8\\
		\hline
		& & & & &&
		\\[-4.3mm]
		$\Lambda_{WB}$ 
		&  0.051 & 0.086 & 0.14 & 0.33 & 0.48 & 1.6&2.3\\
		\hline
		& & & & &&
		\\[-4.3mm]
		$\Lambda_{BB}$ 
		&  0.069 & 0.12 & 0.20 & 0.44 & 0.65 & 2.2&3.1\\
		\hline\hline
		& & & & & &
		\\[-4.3mm]
		$|h_{4,2}|^{}$ & $33$ &$2.0$ & $0.13$ & $1.6\!\times\hsm\!10^{-3}$ &
		$2.0\!\times\hsm\!10^{-4}$ & $3.3\!\times\hsm\!10^{-7}$ & 
		$5.0\!\times\hsm\!10^{-8}$ 
		\\
		\hline
		& & & & & &
		\\[-4.3mm]
		$|h_{3,1}^Z|$ &  $53$ & $6.6$ & $0.83$ & $0.031$
		& $6.6\!\times\hsm\!10^{-3}$ & $5.3\!\times\hsm\!10^{-5}$ & $1.3\!\times\hsm\!10^{-5}$
		\\
		\hline
		& & & & & &
		\\[-4.3mm]
		$|h_{3,1}^\gamma|$
		&  $53$ & $6.6$ & $0.83$ & $0.031$ & $6.6\!\times\hsm\!10^{-3}$
		& $5.3\!\times\hsm\!10^{-5}$& $1.3\!\times\hsm\!10^{-5}$
		\\[0.5mm]
		\hline\hline
	\end{tabular}
\end{center}
\vspace*{-4mm}
\caption{\small{\it 
		Unitarity bounds on the new physics scale $\cut_j^{}$} (in TeV)
	of the dimension-8 nTGC operators and on the nTGC form factors
	$h_j^V$ including both the CPC and CPV cases.\  
	These bounds are derived for various sample values
	of the c.m.\ energy $\,E_{\rm{CM}}^{}$ of the reaction
	$\,q\bar{q}\ito Z\ga\,$ or
	$\,e^-e^+\ito Z\ga\,$ that are relevant to the present collider study.
}
\label{tab:6}
\end{table}
%


With these, we impose the inelastic unitarity condition\,\cite{unitarity} 
on each given partial-wave amplitude, 
$|a_J^{\text{ine}}|\!<\!\frac{1}{\,2\,}$ for $J\!=\!1$,
and derive the following perturbative unitarity bounds
on the leading contributions of the nTGC form factors:
\\[-3mm]
%
\begin{align}
\label{eq:UB-h34}
\hspace*{-2mm}
|h_{4,2}^{}| \!<\! 
\frac{~24\sqrt{2\,}\pi v^2M_Z^2~}{s_W^{}c_W^{}s^2} \hs,
\hspace*{5mm}
|h_{3,1}^Z| \!<\!  
\frac{\,6\sqrt{2\hs}\pi v^2M_Z^{}\,}
{~s_W^{}c_W^{}(T_3^{}\hsm -\hsm Qs_W^2)\,{s^{3/2}\,}}
\hs,\hspace*{5mm}
|h_{3,1}^{\ga}| \!<\! 
\frac{\,6\sqrt{2\hs}\pi v^2M_Z^{}\,}
{~s_W^{2}c_W^{2}|Q|\,{s^{3/2}\,}}
\hs,
\end{align}
%
where $Q$ is the electric charge of the initial state fermions, 
$T_3^{}\!=\!\pm\fr{1}{2}$ for left-handed fermions, and 
$\hs T_3^{}\!=\!0\hs$ for right-handed fermions.\ 
Then, we impose the inelastic unitarity condition on the contribution of
each dimension-8 nTGC operator to the partial wave amplitude 
of the reaction $f\bar{f}\ito Z\ga\hs$,
and derive the following perturbative inelastic unitarity bounds:
\\[-7mm]
\beqs
\label{eq:UB-Lambda8}
\begin{alignat}{3}
&\hspace*{-4mm}
\cut_{G+}^{} \!>\!
\frac{\sqrt{s\,}}{\,(24\sqrt{2\,}\pi )^{1/4}\,}
\,,\quad
&&\cut_{G-}^{} \!>\!
\(\!\!\frac{\,s_W^2|Q|M_Z^{}\,}{\,12\sqrt{2\,}\pi\,}\!\)^{\hsm\!\!\frac{1}{4}}
\!\!(\!\sqrt{s\,})^{\frac{3}{4}} ,
\quad
&&\cut_{\widetilde{B}W}^{} \!>\!
\(\!\!\frac{\,\big|\widehat{T}_3^{}\big| M_Z^{}\,}
{\,12\sqrt{2\,}\pi\,}\!\)^{\hsm\!\!\frac{1}{4}}
\!\!(\!\sqrt{s\,})^{\frac{3}{4}} ,
\\
%
&\hspace*{-4mm}
\cut_{\widetilde{BW}}^{} \!>\!
\(\!\!\frac{\,2\big|Q\big|M_Z^{}\,}
{\,\omega\,}\!\)^{\hsm\!\!\frac{1}{4}}
\!\!(\!\sqrt{s\,})^{\frac{3}{4}} ,
\quad
&&\cut_{\widetilde G+}^{} \!>\!
\frac{\sqrt{s\,}}{~(48\sqrt{2\,}\pi )^{1/4}~}\,,
\quad
&&\cut_{\widetilde G-}^{} \!>\!
\(\!\!\frac{\,s_W^{}|Q|M_Z^{}\,}
{~2\hs c_W^{}\hs\omega\,}\!\)^{\hsm\!\!\frac{1}{4}}
\!\!(\!\sqrt{s\,})^{\frac{3}{4}} ,
\\
&\hspace*{-4mm} 
\cut_{WW}^{} \!>\!
\(\!\!\frac{\,|T_3|M_Z^{}}{\,2\hs \omega\,}\)^{\!\!\frac{1}{4}} 
\!\!(\!\sqrt{s\,})^{\frac{3}{4}}\!,
\quad 
&&\cut_{{BW}}^{} \!>\!
\(\!\!\frac{\,\big|\widehat{Q}\big|M_Z^{}\,}
{\,2s_W^{}c_W^{}\omega\,}\!\!\)^{\hsm\!\!\frac{1}{4}}
\!\!(\!\sqrt{s\,})^{\frac{3}{4}} , 
\quad 
&&\cut_{{BB}}^{} \!>\!
\(\!\!\!\frac{\,2\big|Q \!-\! T_3\big|M_Z^{}\,}
{\,\omega\,}\!\!\)^{\hsm\!\!\frac{1}{4}}
\!\!(\!\sqrt{s\,})^{\frac{3}{4}} \,,
\end{alignat}
\eeqs
where we have defined 
$\,\omega\!=\!12\sqrt{2}\pi/(s_W^{}c_W^{})$,
$\widehat{T}_3^{}\!=\!T_3^{}\!-\!Qs_W^2$, and 
$\widehat{Q}\!=\!Qs_W^2 \!+\! T_3(1\!-\!2s_W^2)$.\ 
With these we derive the numerical unitarity bounds on 
the new physics scale
$\cut_j^{}$ of each nTGC dimension-8 operator and on each nTGC form factor 
$h_j^V$, including both the CPC and CPV cases.\ 
We summarize these unitarity bounds in Table\,\ref{tab:6} 
for a set of sample values of the relevant 
center-of-mass energy $E_{\rm{CM}}^{}\hs (=\!\!\sqrt{s\,}\hs )$ 
of the scattering process $\,f\bar{f}\hsm\ito\hsm Z\ga\,$ 
with $f\bar{f}\!=\!q{\hs}\bar{q}\hs ,\, e^-e^+$.\ 
Table\,\ref{tab:6} demonstrates that these unitarity bounds 
on $\cut_j^{}$ and $h_j^V$ are much weaker than our current collider bounds
(shown in Tables\,I-II of the main text and 
in Tables\,\ref{tab:4}-\ref{tab:5} of this section)
and thus do not affect our collider analyses.


\begin{thebibliography}{}
\bibitem{foot3}
We find that the situation is different for probing the nTGCs at high-energy $e^+e^-$ colliders\,\cite{Ellis:2020ljj}\cite{Ellis:2019zex}\cite{EHXee}, 
where the interference contribution can dominate over the squared contribution.\
\end{thebibliography}
\end{document}